\documentclass[final,3p,times,sort&compress,doublespacing,fleqn]{elsarticle}

\journal{arXiv}

\usepackage{multirow,multicol,setspace}

\usepackage{amsthm,amssymb,array,float,latexsym,amsmath,amssymb,amsbsy,eurosym,theorem,amsfonts,mathrsfs,bbm,verbatim,dsfont,upgreek}
\usepackage{booktabs}
\usepackage{caption}
\usepackage{subcaption}

\usepackage{rotating}
\usepackage{tabularx}
\usepackage[table]{xcolor}
\usepackage[]{graphicx}
\usepackage{color}
\usepackage[most]{tcolorbox}
\usepackage{import}

\usepackage{longtable}

\usepackage{titlesec}
\usepackage{algorithm}
\usepackage{algpseudocode}

\usepackage{scalerel,stackengine}
\newcommand\equalhat{\mathrel{\stackon[1.5pt]{=}{\stretchto{%
    \scalerel*[\widthof{=}]{\wedge}{\rule{1ex}{3ex}}}{0.5ex}}}}

\titleformat{\paragraph}
	{\normalfont\normalsize}{\theparagraph}{1em}{}
\titlespacing*{\paragraph}
	{0pt}{3.25ex plus 1ex minus .2ex}{1.5ex plus .2ex}

\renewcommand{\b}[1]{\boldsymbol{#1}} 
\newcommand{\cc}[1]{\mathcal{#1}}
\newcommand{\s}[1]{\mathscr{#1}}

\renewcommand{\o}[1]{\overline{#1}}

\renewcommand{\t}[1]{\widetilde{#1}}

\usepackage{bm}

\newcommand{\lp}{\left(}
\newcommand{\rp}{\right)}

\newcommand{\lb}{\left[}
\newcommand{\rb}{\right]}

\newcommand{\p}{\partial}

\newcommand{\f}{\displaystyle\frac}

\newcommand{\de}{\delta}

\renewcommand{\d}{\mbox{d}}

\newcommand{\Grad}{\ensuremath{\mbox{Grad}}}

\newcommand{\Div}{\mbox{Div}}

\newcommand{\dyad}{\otimes}

\newcommand{\assembly}[2]{ \mathop{\mbox{\Large {\textsf {\textbf A}}}}_{#1}^{#2} }

\newcommand{\sth}{\{\bullet\}}

\newcommand{\M}{{}^\text{\scriptsize {M}}\!}

\newcommand{\rve}{\footnotesize{RVE}}

\newcommand{\jmp}[1]{[\hspace{-2.0 pt}[ {#1} ]\hspace{-2.0 pt}]}
\newcommand{\avg}[1]{\{\hspace{-2.0 pt}\{ {#1} \}\hspace{-2.0 pt}\}}

\definecolor{gray}{rgb}{0.75, 0.75, 0.75}
\definecolor{yellow}{rgb}{1, 0.7, 0.2}
\definecolor{green}{rgb}{0.3, 0.9, 0.3}
\definecolor{brown}{rgb}{0.6, 0.3, 0.2}
\definecolor{magenta}{rgb}{0.9, 0.1, 0.9}
\definecolor{light}{rgb}{1, 0.7, 0.7}

\graphicspath{{figs/}}

\newcolumntype{Y}{>{\centering\arraybackslash}X}

\begin{document}


\begin{frontmatter}

\title{On the thermal behavior of composites embedding extended general interfaces}

\author[erlangen]{Soheil~Firooz\corref{cor}}
\ead{soheil.firooz@fau.de}
\author[chemnitz]{Kashish~Veda~Eluri}
\author[bilkent]{Ali~Javili}
\author[erlangen,glasgow]{Paul~Steinmann}
\cortext[cor]{Corresponding author.}
\address[erlangen]{Institute of Applied Mechanics, Friedrich--Alexander Universit{\"a}t Erlangen-N{\"u}rnberg, 91058 Erlangen, Germany}
\address[bilkent]{Department of Mechanical Engineering, Bilkent University, 06800 Ankara, Türkiye}
\address[glasgow]{Glasgow Computational Engineering Center, James Watt School of Engineering, University of Glasgow, Glasgow G12 8QQ, United Kingdom}
\address[chemnitz]{Professorship Forming Technology, Chemnitz University of Technology, 09126 Chemnitz, Germany}

\begin{abstract}
In heterogeneous materials, surface roughness, chemical incompatibility, mechanical mismatch or manufacturing defects result in imperfect bonding between the constituents.
Such imperfections introduce a finite-thickness interphase zone between the constituents that possess a graded or inhomogeneous micro-structure.
It is virtually impossible to resolve the behavior of such interphases.
Instead, a well-established approach, is to replace the interphase by a zero-thickness interface endowed with its own energetic structure. 
Conventional interface models define the interface temperature as an average of the temperatures of the surrounding bulk phases. 
However, this assumption is only adequate to represent uniform homogeneous interphases and it fails to accurately capture the behavior of graded or inhomogeneous interphases.
In this work, we address this limitation by allowing for unequal contributions of the surrounding bulk phases to the interface temperature.
We systematically revisit the resulting balance equations and present the computational framework for the finite element implementation of the model.
We further incorporate the proposed model within the homogenization framework and derive closed-form analytical expressions for the effective thermal conductivity of fiber- and particle-reinforced composites, accouting for the influence of the extended general interface.
Finally, through a series of numerical examples, we demonstrate the capabilities of the proposed model and investigate the influence of its parameters on the size-dependent thermal behavior of composites.
\end{abstract}


\end{frontmatter}

\section{Introduction}\label{sec:Introduction}
Over the past few decades, composites have emerged as key engineering materials owing to their exceptional mechanical and thermal properties. 
The overall behavior of a composite is governed not only by the properties of its constituents, but also by microstructural features such as the orientation, distribution, volume fraction, and shape~\cite{Brockenbrough1991,Kouznetsova2001,Chawla2006,Segurado2006,Duan2006,Wang2006a}. 
Predicting the behavior of composites becomes more complex when the bonding between the matrix and the reinforcing phases is imperfect due to surface roughness, chemical incompatibility, mechanical mismatch, or manufacturing defects~\cite{Madhukar1991fiberI,Madhukar1991fiberII,Madhukar1992fiberIII}.
Such imperfections give rise to interphase regions which significantly influence the mechanical and thermal behavior of the composite.

A well-established approach for capturing the influence of a finite interphase is to replace it with a zero-thickness interface characterized by discontinuities in temperature and heat flux~\cite{Sanchez1970,Huy1974}, leading to the notion of \emph{thermal interface models}.
Seminal studies by Hashin~\cite{Hashin1990,Hashin1991c} established the link between interphase properties and interface jumps, highlighting size-dependent effects that become pronounced at small scales.
Thermal interface models can be classified based on the continuity conditions they impose on temperature and heat flux across the interface.
(i) The \emph{Perfect interface model} assumes ideal thermal contact between the constituents, in that both the temperature and the normal heat flux remain continuous across the interface~\cite{Rayleigh1892,Hale1976,hashin1968assessment,miller1969bounds,hamilton1962thermal,han1981effective}.
Analysis of composites with perfect interfaces is, in principle, identical to classical homogenization.
A vast body of literature has been devoted to predicting the effective conductivity of heterogeneous media with perfect interfaces~\cite{Rayleigh1892,Hale1976,hashin1968assessment,miller1969bounds,hamilton1962thermal,han1981effective}.
(ii) The \emph{lowly conductive (LC) interface model} is classically described by Kapitza's model of thermal resistance~\cite{Kapitza1941}. 
This formulation is consistent with the second law of thermodynamics and allows for a discontinuity in temperature across the interface while enforcing continuity of the normal heat flux.
Sanchez-Palencia~\cite{Sanchez1970} demonstrated that a highly resistant interphase of vanishing thickness gives rise to a temperature jump at the interface.
Several micromechanical approaches were developed to quantify the impact of LC interfaces on the effective thermal properties of composites. 
Benveniste and co-workers~\cite{Benveniste1986,Benveniste1987a} generalized classical homogenization schemes, including the generalized self-consistent method and the Mori--Tanaka theory, to account for LC interfaces and showed that temperature jumps systematically reduce the effective conductivity. 
Further advances were made through both analytical and variational methods. 
Chiew and Glandt~\cite{Chiew1987}, Hasselman and Johnson~\cite{Hasselman1987}, Hasselman et al.~\cite{hasselman1993effective}, and Every et al.~\cite{Every1992} incorporated interfacial resistance into modified classical models and linked size effects to Kapitza-type barriers.
See~\cite{quang2011eshelby,DuanKarihaloo2007,quang2011f,Kushch2014,Javili2012,Bonfoh2017} for further studies on LC interface model.
(iii) The \emph{highly conductive (HC) interface model} is characterized by continuous temperature across the interface and a discontinuity in the normal heat flux, arising from in-plane heat conduction along the interface.
The model was first formalized by Pham-Huy and Sanchez-Palencia~\cite{Huy1974}, who showed that for a superconducting interphase of vanishing thickness, the flux jump is proportional to the surface Laplacian of the interfacial potential.
Subsequent studies established theoretical bounds and dualities between HC and LC interfaces~\cite{Lipton1997,Lipton1997a,Cheng1997a,Lipton1998}.
Miloh and Benveniste~\cite{Miloh1999} and Chen~\cite{Chen2001a} extended HC interface model for ellipsoidal and circular inclusions with spatially varying interface properties, while Duan and Karihaloo~\cite{Duan2007} unified HC and LC descriptions through a multiple-scattering framework.
Le-Quang et al.~\cite{Le-Quang2010} extended Eshelby's solution to anisotropic phases with HC interfaces, showing uniformity of the thermal tensor fields in contrast to earlier studies. 
See~\cite{bonfoh2018,Le-Quang2010,le2011effective,Javili2013f} for further studies on HC interface model.
(iv) The \emph{general interface model} allows for discontinuities in both temperature and the normal heat flux across the interface.
This broader formulation includes the lowly conducting and highly conducting interfaces as special cases.
Seminal work of Hashin~\cite{Hashin2001} established the theoretical foundation of this interface model in the context of elasticity.
Variational estimates and bounds for composites with imperfect interfaces were developed by Lipton and Talbot~\cite{lipton2001bounds}.
These works was followed by important theoretical refinements.
Gu and co-workers~\cite{Gu2011,Gu2011a,Gu2015} developed coordinate-free derivations and closed-form estimates for the size-dependent effective conductivity of composites with spherical inclusions.
These models have since been embedded in homogenization schemes such as Mori--Tanaka and self-consistent methods, and extended to anisotropic and spheroidal composites~\cite{LeQuang2013,le2016,kushch2015,levin2016,koutsawa2018micromechanics}.
Higher-order asymptotic analyses further refined the modeling of curved and finite-thickness interphases, leading to interface conditions of arbitrary order accuracy~\cite{baranova2020,han2021,serpilli2019,kushch2024higher,serpilli2021,serpilli2017asymptotic,dumont2018imperfect,han2019consistent}.
A thermodynamically consistent formulation of general imperfect interfaces, allowing simultaneous jumps in temperature and normal heat flux, was developed by Javili et al.~\cite{javili2014general}, together with a unified finite-element framework.
Kaessmair et al.~\cite{kaessmair2014thermomechanics} extended this formulation to thermomechanically coupled coherent interfaces.
For further computational approaches on thermal interface problems, see~\cite{han2019consistent,Yvonnet2008,Yvonnet2011a}.

\begin{figure}[h!]
	\centering
	{\includegraphics[width=1.0\textwidth]{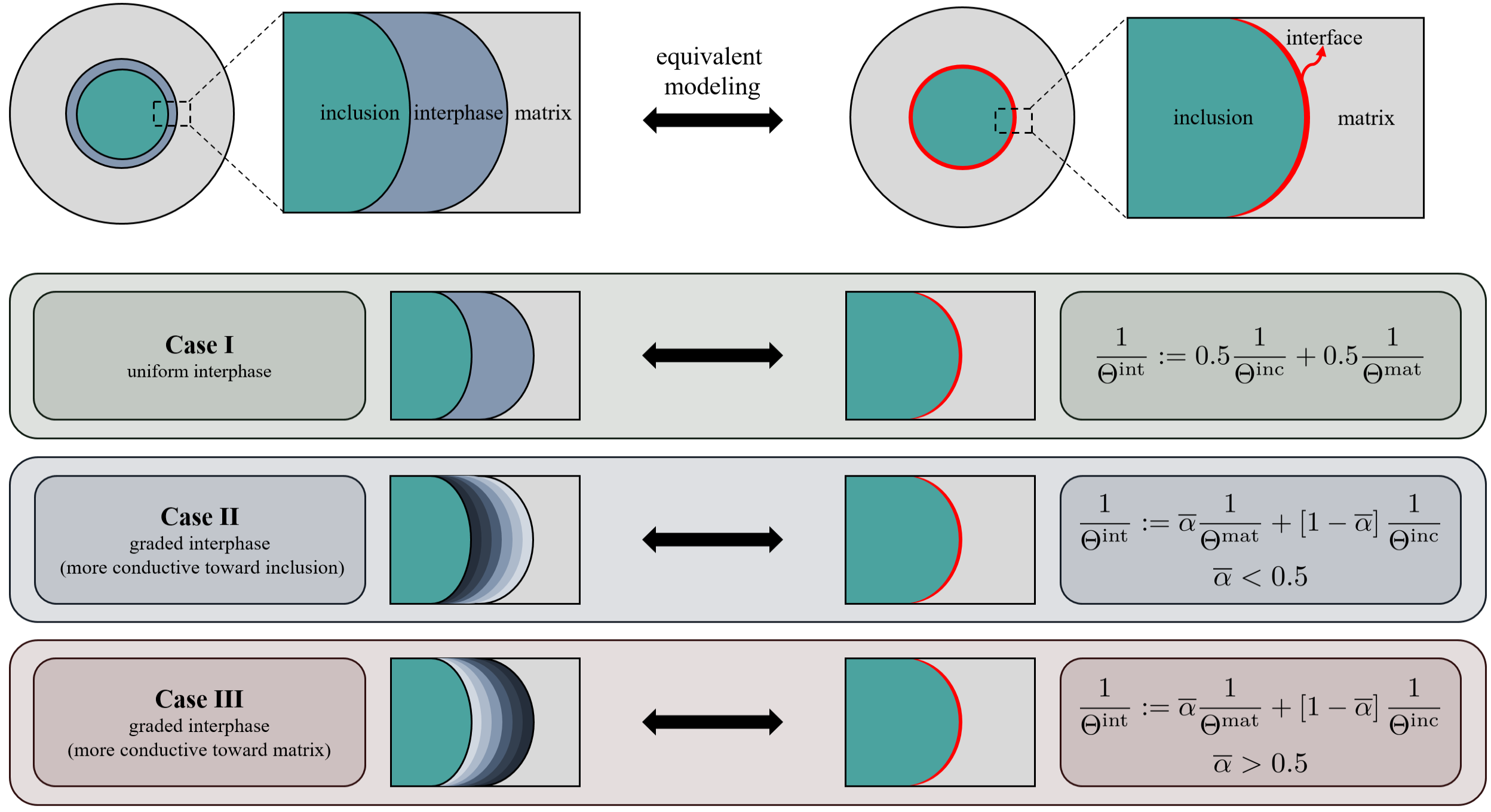}}
	\caption{
	Schematic illustration of a finite-thickness interphase and its equivalent zero-thickness interface model. Temperature is denoted by $\Theta$, and coldness is defined as $1/\Theta$. The interfacial coldness is expressed as a weighted average of the coldness on the inclusion and matrix sides, with $\o{\alpha}$ denoting the weighting factor. For a uniform interphase (Case I), both sides contribute equally. For a graded interphase with conductivity increasing toward the inclusion (Case II), the inclusion side contributes more strongly, whereas for conductivity increasing toward the matrix (Case III), the matrix side has a greater contribution.
	}
	\label{fig:Mot}
\end{figure}

To calculate the interface field variables, such as the interface temperature, the \emph{classical general thermal interface model} assumes that the surrounding bulk domains, namely the inclusion and the matrix, contribute equally.
This assumption is, however, simplifying and is sufficient only to capture the influence of a uniform interphase.
For graded interphases, which consist of layers with different conductivities, the general interface model fails, and a more generic model is required that enables unequal contributions from the surrounding materials to the interface quantities.
This concept is schematically illustrated in Fig.~\ref{fig:Mot}.
We consider three different interphase structures and model each of them with an equivalent interface.
Only for Case I, which has a uniform interphase, does the equal contribution assumption provide accurate results.
For interphases with varying conductivity as shown in cases II and III, the interface tends to receive more contribution from one side than the other.\footnote{The example in Fig.~\ref{fig:Mot} is given to motivate the need for the weighting parameter $\o{\alpha}$ in capturing the asymmetric contributions from the surrounding bulk domains.
More details on this example and precise values of the weighting parameter $\o{\alpha}$ are provided in Sec.~\ref{sec:interface_vs_interphase}.
Here, $\o{\alpha}$ represents the relative contribution from the inclusion side, with $1-\o{\alpha}$ representing the contribution from the matrix side.
The relationships between the interfacial coldness and the coldness on the inclusion and matrix sides are obtained from thermodynamic considerations.
}
We introduce the \emph{extended general thermal interface model}, which generalizes the classical model by allowing unequal contributions from the surrounding bulk domains to the interface field variables via a weighting parameter $\o{\alpha}$.
The necessity of such a generalization for mechanical problems only was demonstrated by Saeb et al.~\cite{Saeb2021}, with related developments presented in~\cite{Firooz2021,Firooz2022}.
However, the extension of this concept to thermal problems has not been explored, which motivates the present study.
More importantly, the thermal problem leads to a better understanding of the theory in that we are able to identify the value of $\o{\alpha}$ via analytical or numerical means for a given interphase structure.
To this end, the governing balance equations and thermodynamic restrictions are revisited and systematically re-derived though weighted averaging. 
The resulting formulation encompasses the classical interface models as special cases while providing a more generic description of graded and nonuniform interphases, thereby improving the fidelity of thermal interface modeling.
In summary, the key objectives of this work are:
\begin{itemize}
\item To develop an extended general interface model with unequal inclusion and matrix contributions to interface quantities and systematically reformulate the associated continuum balance equations;
\item To establish finite element and interface-enhanced homogenization frameworks, including extended composite assemblage methods, linking microscopic interface effects to the effective thermal conductivity of fiber- and particle-reinforced composites;
\item To assess the model's significance and utility, and parameter effects on size-dependent effective thermal properties, with numerical examples and computational simulations.
\end{itemize}

The remainder of this manuscript is organized as follows.
Section~2 provides the governing equations of a continuum body embedding an extended general interface.
Section~3 elaborates the finite element implementation of the proposed interface model.
Section~4 presents the homogenization framework used to define the overall thermal conductivity of composites.
Section~5 develops analytical methods to estimate the overall conductivity of fiber- and particle-reinforced composites.
Section~6 presents numerical examples and finally Section~7 concludes the manuscript with a summary of the main findings.


\section{Governing equations}\label{sec:Governing equations}

This section provides the governing equations of a continuum body embedding an extended general interface.
We begin with some preliminaries, introducing the problem definition along with the relevant control volume and domain definitions, and then proceed with the balance equations.
In this work, we restrict our attention to steady-state thermal problems.
For related works on surface elasticity and energetic interface formulations in mechanical problems, the reader is referred to~\cite{Fried2007,Fried2005,Sharma2004,Sharma2004a,Sharma2007a,Steigmann1999}.

\subsection{Preliminaries}

\begin{figure}[h!]
	\centering
	{\includegraphics[width=1.0\textwidth]{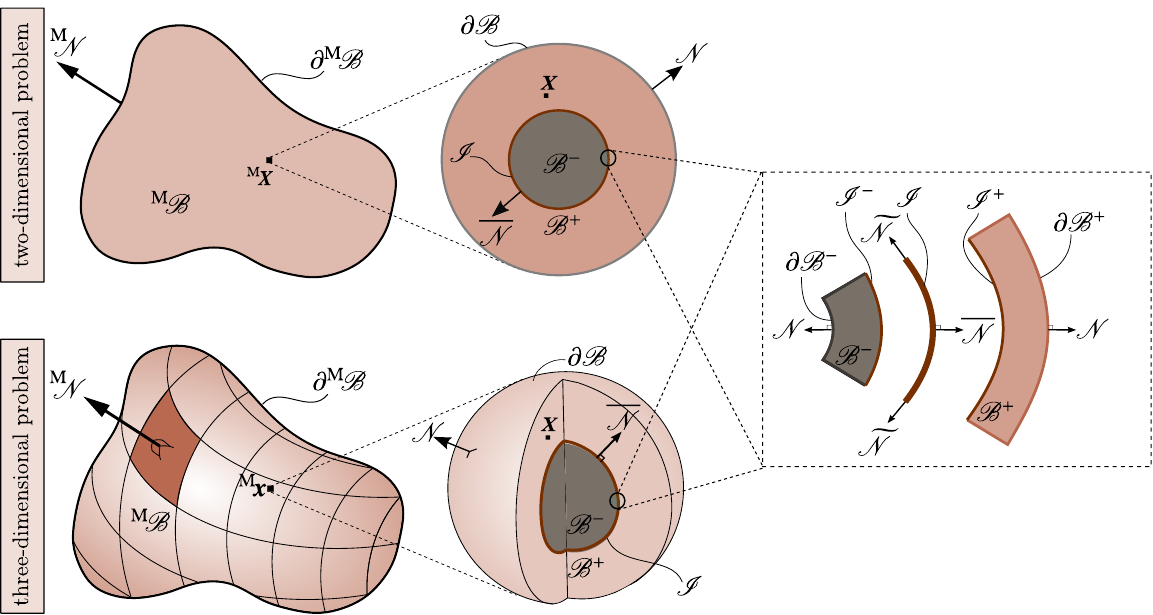}}
	\caption
	{
	Schematic representation of a heterogeneous continuum body at the macroscale, $\M{\s{B}}$, together with its representative volume element (RVE) at the microscale, $\s{B}$.  
	At the microscale, the bulk is divided into two subdomains, $\s{B}^{+}$ and $\s{B}^{-}$, by a zero-thickness interface $\s{I}$. 
	The outward unit normal to the body boundary is $\s{N}$.
	The unit normal to the interface is $\o{\s{N}}$ pointing from the minus side to the plus side, and the unit normal vector to the boundary of the interface, but tangential to the interface, is denoted $\t{\s{N}}$.
	}
	\label{fig:problem}
\end{figure}

Consider a continuum body representing a heterogeneous medium that occupies the configuration $\M{\s{B}}$ with the boundary $\p\M{\s{B}}$ at the macroscale, as shown in Fig.~\ref{fig:problem}.
The corresponding representative volume element ({\rve}) at the microscale occupies the configuration $\s{B}$ with the boundary $\p\s{B}$.
In order to derive analytical solutions for effective isotropic response, the {\rve} is assumed to be circular in 2D and spherical in 3D, representing fiber- and particle-reinforced composites, respectively.
At the microscale, the bulk is partitioned into two disjoint subdomains, $\s{B}^{+}$ and $\s{B}^{-}$, by a zero-thickness interface $\s{I}$ such that $\s{B} = \s{B}^{+}\cup\s{B}^{-}\cup\s{I}$.
The intersection of the interface with its plus side and minus side are denoted as $\s{I}^{+} = \s{I} \cap \s{B}^{+}$ and $\s{I}^{-} = \s{I} \cap \s{B}^{-}$, respectively.
The outward unit normal to the boundary of the body at the microscale is $\s{N}$, the unit normal to the interface is $\o{\s{N}}$ pointing from the minus to the plus side, and the unit normal to the boundary of the interface but tangential to the interface is denoted as $\t{\s{N}}$.

\begin{figure}[h!]
	\centering
	{\includegraphics[width=0.8\textwidth]{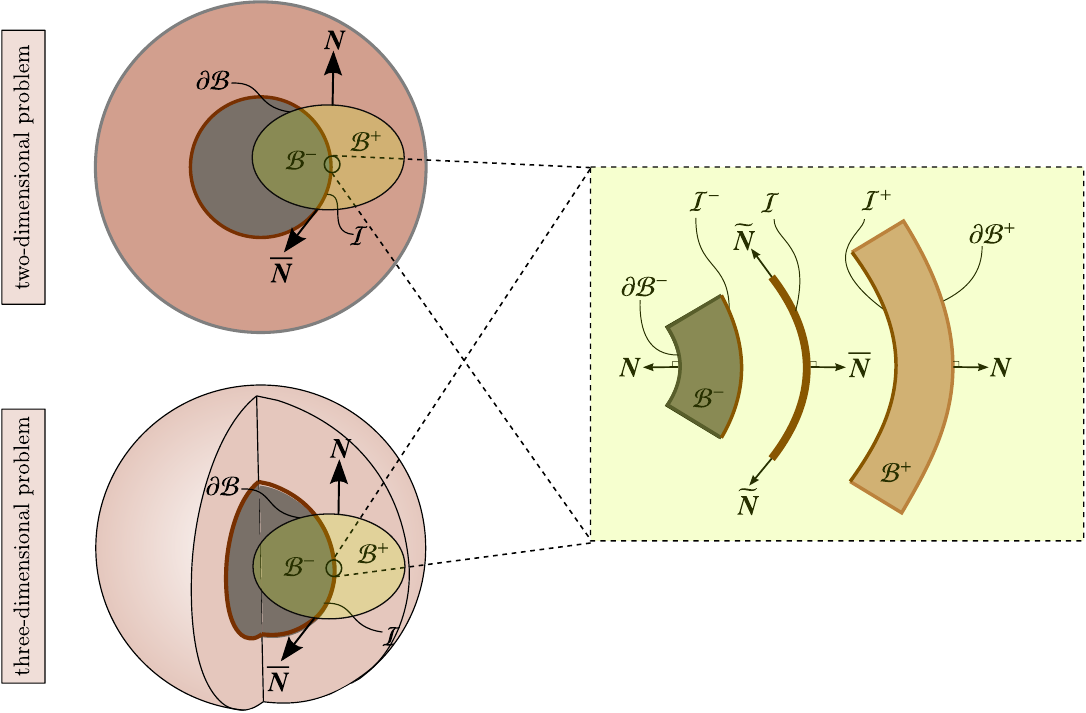}}
	\caption{
	Control region $\cc{B}$ within the RVE, with boundary $\p\cc{B}$ and outward normal $\b{N}$. 
	The control region may intersect the interface $\s{I}$, giving the sub-interface $\cc{I}$, which divides $\cc{B}$ into subdomains $\cc{B}^+$ and $\cc{B}^-$. 
	The interface sides $\cc{I}^+$ and $\cc{I}^-$ are also indicated together with their respective unit normal vectors.
	}
	\label{fig:controlregion}
\end{figure}

The governing equations for thermal problems in various regions of the {\rve} are derived via applying the energy and entropy equations over an arbitrary control region in integral forms, and then localization to a point within the bulk $\s{B}$ and on the interface $\s{I}$.
The canonical control region for the {\rve}, as shown in Fig.~\ref{fig:controlregion}, is defined as $\cc{B} \subset \s{B}$ with boundary $\p{\cc{B}}$.
The outward unit normal to $\p{\cc{B}}$ is denoted as $\b{N}$.
The control region can contain a section of the interface $\s{I}$, denoted as $\cc{I}= \cc{B} \cap \s{I}$. 
Note that the subdomain $\cc{B}$ may be fully enclosed within the inclusion or the matrix, or it may be chosen such that it includes the interface corresponding to $\p{\cc{B}} \cap \cc{I} = \emptyset$ or $\p{\cc{B}} \cap \cc{I} \neq \emptyset$.
We denote the boundary of the interface as $\p\cc{I}$ with outward unit normal $\o{\b{N}}$.
The interface $\cc{I}$ divides the control region $\cc{B} $ into two subdomains, $\cc{B}^{+}$ and $\cc{B}^{-}$. 
The interface $\cc{I}$ consists of two sides $\cc{I}^+ = \cc{B}^+ \cap \cc{I}$ and $\cc{I}^- = \cc{B}^- \cap \cc{I}$, with $\cc{I} = \cc{I}^+ \cup \cc{I}^-$.
The bulk and interface temperatures are denoted $\Theta$, and $\o{\Theta}$, respectively.
The temperature jump across the interface is
\begin{equation}
\jmp{\Theta} = \Theta^{+} - \Theta^{-}\,,
\label{eq:T-jump}
\end{equation}
with $\Theta^{+}$ and $\Theta^{-}$ being the temperatures on the plus and minus sides of the interface, respectively.
Unlike the classical interface models, \emph{we do not impose the interface temperature as the average of the temperature on its sides}, since this simplifying assumption can lead to inaccurate results when comparing the zero-thickness interface model with its associated finite-thickness interphases~\cite{Saeb2021}.
Here, we propose an \emph{extended general interface model} in which the interface \emph{coldness}, i.e., the inverse interface temperature $1/\o{\Theta}$ entering the balance equations, is defined as a weighted average of the bulk coldnesses across the interface, allowing for unequal contributions from the plus and minus sides.
That is
\begin{equation}
\boxed{
\f{1}{\o{\Theta}}=\avg{\f{1}{\Theta}}_{\o{\alpha}} \qquad \text{and} \qquad \avg{\f{1}{\Theta}}_{\o{\alpha}} := \o{\alpha} \f{1}{\Theta^{+}}  + \lb 1 - \o{\alpha} \rb \f{1}{\Theta^{-}} \qquad \text{with} \qquad 0 \leq \o{\alpha} \leq 1 \,.
}
\label{eq:avg}
\end{equation}
Based on this definition, the interface temperature then reads
\begin{equation}
\o{\Theta}
=\lb \avg{\f{1}{\Theta}}_{\o{\alpha}} \rb^{-1}\!\!\!
=\lb \o{\alpha}\frac{1}{\Theta^{+}} + \lb1-\o{\alpha}\rb\frac{1}{\Theta^{-}} \rb^{-1}\!\!\!
=\lb\f{\o{\alpha}\,\Theta^{-} + \lb1-\o{\alpha}\rb\Theta^{+}}{\Theta^{+}\Theta^{-}}\rb^{-1}\!\!\!
=\f{\Theta^{+}\Theta^{-}}{\o{\alpha}\Theta^{-}+\lb1-\o{\alpha}\rb\Theta^{+}}
=\f{\Theta^{+}\Theta^{-}}{\avg{\Theta}_{[1-\o{\alpha}]}}\,.
\label{eq:Theta-avg}
\end{equation}
For $\o{\alpha}=0.5$, Eq.~\eqref{eq:avg} gives the arithmetic mean of the bulk coldnesses, while the interface temperature is the harmonic mean of the temperatures on the plus and minus sides,
\begin{equation}
\f{1}{\o{\Theta}}
=\f{1}{2}\lb\f{1}{\Theta^{+}}+\f{1}{\Theta^{-}}\rb
\qquad \Leftrightarrow \qquad
\o{\Theta}
=\f{2\Theta^{+}\Theta^{-}}{\Theta^{+}+\Theta^{-}}\,.
\end{equation}
Next, for the sake of presentation, we identify the temperature gradients as ``thermal intensity'' in the bulk and on the interface as $\b{E} := \Grad\,\Theta$ and $\o{\b{E}} := \o{\Grad}\,\o{\Theta}$, respectively.
The interface gradient is defined by $\o{\Grad} \sth := \Grad \sth \cdot \o{\b{I}}$ with the interface identity tensor $\o{\b{I}} := \b{I} - \o{\b{N}}\dyad\o{\b{N}}$.
The interface divergence operator is the trace of the interface gradient.

\subsection{Balance equations}

Equipped with the definitions and fundamental concepts, we proceed to elaborate on the balance equations of a steady-state thermal problem accounting for interfaces.
The balance equations correspond to the microscale, and therefore the source terms are omitted.
The global heat balance for bulk and interface reads
\begin{equation}
\begin{aligned}
&\int_{\p \cc{B}} [-\b{Q} \cdot \b{N}] \, \,\d A = 0 
\qquad &&\text{for}\,\,  &&\p{\cc{B}} \cap \cc{I} = \emptyset \,,\\
&\int_{\p \cc{B}/\cc{I}} [-\b{Q} \cdot \b{N}] \,\d A + \int_{\p \cc{I}} [-\o{\b{Q}} \cdot \t{\b{N}}] \,\d L = 0
\qquad &&\text{for}\,\,  &&\p{\cc{B}} \cap \cc{I} \neq \emptyset\,,
\end{aligned}
\label{eq:heat-balance-global}
\end{equation}
where $\b{Q}$ and $\o{\b{Q}}$ are the bulk heat flux per unit area and the interface heat flux per unit length, respectively.
Taking into account the arbitrariness of the control region $\cc{B}$, Eq.~\eqref{eq:heat-balance-global} can be written in local form as
\begin{equation}
\boxed{
\begin{aligned}
&\Div\,\b{Q}= 0
\qquad &&\text{in}\,\,\s{B} \,,\\
&\o{\Div}\,\o{\b{Q}} + \jmp{\b{Q}}\cdot \o{\b{N}} = 0
\qquad &&\text{on}\,\,\s{I}\,.
\end{aligned}
}
\label{eq:heat-balance-local}
\end{equation}
In a similar fashion, the bulk and interface internal entropy balance equations are introduced next.
Let $\b{H}$ and $\o{\b{H}}$ be the bulk entropy flux per unit area and the interface entropy flux per unit length, respectively.
Furthermore, we define $\Pi \geq 0 $ and $\o{\Pi} \geq 0 $ as the (positive) bulk entropy production rate per unit volume and interface (positive) entropy production rate per unit area, respectively.
The integral forms of entropy balance equations read
\begin{equation}
\begin{aligned}
& \int_{\p \cc{B}} [-\b{H} \cdot \b{N}]  \,\d A + \int_{\cc{B}} \Pi \,\d V = 0 
\qquad &&\text{for}\,\, &&\p{\cc{B}} \cap \cc{I} = \emptyset \,,\\
&\int_{\p \cc{B}/\cc{I}}\!\!\!\! [-\b{H} \cdot \b{N}] \,\d A + \int_{\cc{I}} \o{\Pi} \,\d A   + \int_{\p \cc{I}} [-\o{\b{H}} \cdot \t{\b{N}}] \,\d L = 0 
\qquad &&\text{for}\,\, &&\p{\cc{B}} \cap \cc{I} \neq \emptyset \,.
\end{aligned}
\label{eq:entropy-balance-global}
\end{equation}
Taking into account the arbitrariness of the control region $\cc{B}$, Eq.~\eqref{eq:entropy-balance-global} can be written in local form as
\begin{equation}
\begin{aligned}
&\Div\,\b{H} - \Pi  = 0
\qquad &&\text{in}\,\,\s{B} \,,\\
&\o{\Div}\, \o{\b{H}}-\o{\Pi} + \jmp{\b{H}}\cdot \o{\b{N}} = 0
\qquad &&\text{on}\,\,\s{I}\,.
\end{aligned}
\label{eq:entropy-balance-local}
\end{equation}
Using a Coleman–Noll-like procedure, we assume that the entropy fluxes are the corresponding heat fluxes divided by the corresponding absolute temperatures as ${\b{H}}=\b{Q}/\Theta$ and $\o{\b{H}}=\o{\b{Q}}/\o\Theta$.
Furthermore, the dissipation in the bulk and on the interface read $\cc{D} := \Pi\, \Theta$ and $\o{\cc{D}} := \o{\Pi}\, \o{\Theta}$, respectively.
Therefore, the entropy balance relations~\eqref{eq:entropy-balance-local} read as
\begin{equation}
\begin{aligned}
& \Pi  = \Div\,\b{H} && \quad \Longrightarrow \quad && \cc{D} = \Theta \, \Div\,\f{\b{Q}}{\Theta}
\qquad &&\text{in}\,\s{B} \,,\\
& \o{\Pi} = \o{\Div}\, \o{\b{H}} + \jmp{\b{H}}\cdot \o{\b{N}} = 0 && \quad \Longrightarrow \quad && \o{\cc{D}} = \o{\Theta} \, \o{\Div}\, \f{\o{\b{Q}}}{\o\Theta} + {\o\Theta}\jmp{\f{\b{Q}}{\Theta}}\cdot \o{\b{N}}
\qquad &&\text{on}\,\s{I}\,.
\end{aligned}
\label{eq:entropy-balance-local2}
\end{equation}
Using the relation
\begin{equation}
\jmp{\f{\b{Q}}{\Theta}} = \jmp{\f{1}{\Theta}}\avg{\b{Q}}_{[1-\o{\alpha}]} + \avg{\f{1}{\Theta}}_{\o{\alpha}}\jmp{\b{Q}}\,,
\end{equation}
the dissipation equations are expanded to
\begin{equation}
\begin{aligned}
& \cc{D} = \Div\,{\b{Q}} + \Theta \, \b{Q} \cdot \big[ \Grad \,{\Theta}^{-1} \big]
\qquad &&\text{in}\,\,\s{B} \,,\\
& \o{\cc{D}} = \o{\Div}\, \o{\b{Q}} + \o{\Theta} \, \o{\b{Q}} \cdot \, \big[ \o\Grad \, \o{\Theta}^{-1} \big]  + {\o\Theta}\jmp{\f{1}{\Theta}}\avg{\b{Q}}_{[1-\o{\alpha}]} \cdot \o{\b{N}} + {\o\Theta}\avg{\f{1}{\Theta}}_{\o{\alpha}}\jmp{\b{Q}} \cdot \o{\b{N}} \qquad &&\text{on}\,\,\s{I}\,.
\end{aligned}
\label{eq:entropy-balance-local3}
\end{equation}
Incorporating the heat balance equations~(\ref{eq:heat-balance-local}), and using the identities
\begin{equation}
\begin{aligned}
	\Grad \,\f{1}{\Theta} = - \f{1}{\Theta^2} \, \Grad \,{\Theta} \quad , 
	\quad \Grad \, \ln \Theta = \f{1}{\Theta} \, \Grad \,{\Theta} \quad \Rightarrow \quad \Theta \, \b{Q} \cdot \big[ \Grad \,{\Theta}^{-1} \big] = -\b{Q}\cdot \text{Grad}\,{\ln}\,\Theta \,,\\
	\Grad \,\f{1}{\o\Theta} = - \f{1}{\o\Theta^2} \, \Grad \,{\o\Theta} \quad , 
	\quad \Grad \, \ln \o\Theta = \f{1}{\o\Theta} \, \Grad \,{\o\Theta} \quad \Rightarrow \quad \o{\Theta} \, \o{\b{Q}} \cdot \, \big[ \o\Grad \, \o{\Theta}^{-1} \big] = -\o{\b{Q}}\cdot \o{\text{Grad}}\,{\ln}\,\o{\Theta} \,,\\
\end{aligned}
\end{equation}
the dissipation inequality for the bulk and interface eventually can be expressed as
\begin{equation}
\boxed{
\begin{aligned}
&\cc{D} =-\b{Q}\cdot \text{Grad}\,{\ln}\,\Theta \geq 0
\qquad &&\text{in}\,\,\s{B} \,,\\
&\o{\cc{D}}=
-\o{\b{Q}}\cdot \o{\text{Grad}}\,{\ln}\,\o{\Theta} 
+
{\o\Theta}\jmp{\f{1}{\Theta}}\avg{\b{Q}}_{[1-\o{\alpha}]}\cdot\o{\b{N}}
+
{\o\Theta}\left[\avg{\f{1}{\Theta}}_{\o{\alpha}}-\f{1}{\o{\Theta}}\right]\jmp{\b{Q}} \cdot \o{\b{N}} \geq 0
\qquad &&\text{on}\,\,\s{I}\,.
\label{eq:dissipation}
\end{aligned}
}
\end{equation}
The interface dissipation admits an additive decomposition into three contributions as
\begin{equation}
\boxed{
\begin{aligned}
&\o{\cc{D}}_{\text{B}}=-\o{\b{Q}}\cdot \o{\text{Grad}}\,{\ln}\,\o{\Theta} \geq 0 
&& \text{membrane-associated dissipation on the interface} \,,\\
&\o{\cc{D}}_{\text{C}}={\o\Theta}\jmp{\f{1}{\Theta}}\avg{\b{Q}}_{[1-\o{\alpha}]}\cdot\o{\b{N}}\geq 0 && \text{cohesion-associated dissipation on the interface} \,,\\
&\o{\cc{D}}_{\text{D}}={\o\Theta}\left[\avg{\f{1}{\Theta}}_{\o{\alpha}}-\f{1}{\o{\Theta}}\right]\jmp{\b{Q}} \cdot \o{\b{N}}\geq 0 && \text{deviation-associated dissipation on the interface} \,,
\label{eq:dissipation-int}
\end{aligned}
}
\end{equation}
which can sufficiently fulfill the interface dissipation inequality, following the terminology of \citep{Javili2025}.
To satisfy the bulk dissipation inequality $\cc{D}\geq0$ and the interface membrane-associated dissipation inequality $\o{\cc{D}}_{\text{B}} \geq 0$, we use Fourier's law which suggests
\begin{equation}
\boxed{
\begin{aligned}
&\b{Q} =-\b{K}\cdot \text{Grad}\,\Theta  \qquad &&\text{in}\,\s{B}\,, \\
&\o{\b{Q}} =-\o{\b{K}}\cdot\o{\text{Grad}}\,\o{\Theta} \qquad &&\text{on}\,\s{I}\,, \\
\end{aligned}
}
\label{eq:Fourier}
\end{equation}
where $\b{K}$ and $\o{\b{K}}$ denote the positive semi-definite thermal conductivity tensors for the bulk and the interface, respectively.
Due to the identities
\begin{equation}
\begin{aligned}
	\jmp{\f{1}{\Theta}} = - \f{\avg{\Theta^{-1}}}{\avg{\Theta}} \jmp{\Theta} \qquad \text{and} \qquad \f{\avg{\Theta^{-1}}}{\avg{\Theta}} > 0 \,,
\end{aligned}
\end{equation}
the interface cohesion-associated dissipation can be alternatively expressed as
\begin{equation}
\begin{aligned}
\o{\cc{D}}_{\text{C}}={\o\Theta}\jmp{\f{1}{\Theta}}\avg{\b{Q}}_{[1-\o{\alpha}]}\cdot\o{\b{N}}\geq 0  \qquad \Rightarrow \qquad - \jmp{\Theta} \avg{\b{Q}}_{[1-\o{\alpha}]}\cdot\o{\b{N}}\geq 0 \,,
\end{aligned}
\end{equation}
which can be satisfied via imposing the constitutive constraint
\begin{equation}
\boxed{
\begin{aligned}
 \avg{\b{Q}}_{[1-\o{\alpha}]}\cdot\o{\b{N}} = -\f{1}{\o{m}} \jmp{\Theta} \qquad \text{with} \qquad \o{m} \geq 0 \,,
\end{aligned}
}
\end{equation}
in which $\o{m}$ is the well-known Kapitza heat resistance coefficient.
Finally, the weighted average definition~(\ref{eq:avg}) for the interface temperature trivially fulfills the remaining condition $\o{\cc{D}}_{\text{D}} \geq 0$ to ensure thermodynamics consistency.\footnote{In the absence of the temperature definitions~(\ref{eq:avg}),
one can impose ${\o{\Theta}}{}^{-1}-\avg{\Theta^{-1}}_{\o{\alpha}} = -\o{s}\,\jmp{\b{Q}} \cdot \o{\b{N}}$ 
with a heat flux sensitivity coefficient $\o{s} \geq 0$.
For the case of $\o{s}=0$, the interface temperature definition reduces to the weighted average form~(\ref{eq:avg}).
}

\section{Finite element implementation}\label{sec:FEM}

The aim of this section is to develop a finite element framework capable of incorporating the extended general interface model. 
To this end, a curvilinear-coordinate-based finite element method~\cite{Javili2014} is employed, as it naturally reflects the underlying geometrical and mathematical concepts of one-dimensional manifolds embedded in a two-dimensional space and two-dimensional manifolds embedded in a three-dimensional space.
For comprehensive discussions of the finite element formulation and its implementation, the reader is referred to~\cite{Wriggers2008,Hughes2012}, among others.
The primary step towards the finite element implementation of the theory is to derive the weak form of the governing equations.
For this purpose, the strong form of the temperature evolution equation in the bulk and on the interface is multiplied from the left by the scalar-valued test functions $\delta\Theta$ and $\delta\o{\Theta}$, respectively, and are integrated over the bulk and interface domain in the material configuration. 
Note that $\delta\Theta \in \cc{H}^{1}(\cc{B})$ and $\delta\o{\Theta} \in \cc{H}^{1}(\cc{I})$ where $\cc{H}^{1}$ denotes the Sobolev space of order one.
Using the extended divergence theorem, the weak form on the bulk reads 
\begin{equation}
\begin{aligned}
&
\int_{\s{B}}  \Div \b{Q} \,  \delta \Theta \, \d V  =\\
-
&\int_{\s{B}} \b{Q} \cdot \Grad \delta \Theta \, \d V
+
\int_{\s{B}} \Div [\delta \Theta \, \b{Q}] \, \d V =\\ 
-
&\int_{\s{B}} \b{Q} \cdot \Grad \delta \Theta \, \d V
+
\int_{\p\s{B}} \lb  \b{Q}\cdot \b{N} \rb  \, \delta \Theta \, \d A 
-
\int_{\s{I}} \jmp{ \delta \Theta \, \b{Q} }  \cdot \o{\b{N}} \, \d A=\\
-
&\int_{\s{B}} \b{Q} \cdot \Grad \delta \Theta \, \d V
+
\int_{\p\s{B}} \lb  \b{Q}\cdot \b{N} \rb  \, \delta \Theta \, \d A 
-
\int_{\s{I}} \avg{\delta \Theta}_{\o{\alpha}}\, \jmp{\b{Q}}  \cdot \o{\b{N}} \, \d A 
-
\int_{\s{I}} \jmp{\delta \Theta}\, \avg{\b{Q}}_{\lb1-\o{\alpha}\rb}  \cdot \o{\b{N}} \, \d A =\\
-
&\int_{\s{B}} \b{Q} \cdot \Grad \delta \Theta \, \d V
+
\int_{\p\s{B}} \underbrace{\lb  \b{Q}\cdot \b{N} \rb}_{Q^{\text{p}}}  \, \delta \Theta \, \d A 
-
\int_{\s{I}} \avg{\delta \Theta}_{\o{\alpha}}\, \jmp{\b{Q}}  \cdot \o{\b{N}} \, \d A 
+
\int_{\s{I}} \f{1}{\o{m}} \jmp{\Theta}\, \jmp{\delta \Theta}\, \d A =0\,,
\end{aligned}
\label{eq:FEM-bulk}
\end{equation}
with ${Q^{\text{p}}}$ denoting the prescribed normal heat flux on the Neumann part of the external boundary.
Similarly, the interface weak from reads
\begin{equation}
\begin{aligned}
&\int_{\s{I}}  \o\Div\,\o{\b{Q}} \, {\delta \o{ \Theta}} \, \d A +
\int_{\s{I}}  \lb \jmp{\b{Q}} \cdot \o{\b{N}} \rb \delta\o{ \Theta } \, \d A =\\
-
&\int_{\s{I}} \o{\b{Q}}\cdot\o{\Grad}\,{\delta \o{ \Theta}} \, \d A
+
\int_{\s{I}} \o{\Div} [\delta \o{\Theta} \, \o{\b{Q}}] \, \d A
+
\int_{\s{I}} \lb \jmp{\b{Q}} \cdot \o{\b{N}} \rb \, \delta\o{ \Theta } \, \d A =\\
-
&\int_{\s{I}} \o{\b{Q}}\cdot\o{\Grad}\,{\delta \o{ \Theta}} \, \d A
+
\int_{\s{I}}    \o{\cc{C}} \lb  \o{\b{Q}} \cdot \o{\b{N}} \rb     \delta \o{\Theta}  \, \d A
+
\int_{\p\s{I}}  \lb  \o{\b{Q}} \cdot \t{\b{N}} \rb  \delta \o{\Theta}  \, \d L 
+
\int_{\s{I}} \lb \jmp{\b{Q}} \cdot \o{\b{N}} \rb \, \delta\o{ \Theta } \, \d A = 0\,,
\end{aligned}
\label{eq:FEM-interface}
\end{equation}
with $\o{\cc{C}}$ representing the interface curvature.
The seond term in Eq.~\eqref{eq:FEM-interface} vanishes owing to the superficiality of the interface and the third term vanishes since the interface in our composite medium is assumed to be closed.
Finally, through combining Eqs.~\eqref{eq:FEM-bulk} and~\eqref{eq:FEM-interface}, the resulting weak form reads
\begin{equation}
\boxed{
\begin{aligned}
\int_{\s{B}}\!\! \b{Q} \cdot \Grad \delta \Theta \, \d V
\!+\!
\int_{\s{I}}\!\! \o{\b{Q}}\cdot\o{\Grad}\,{\delta \o{ \Theta}} \, \d A
\!+\!
\int_{\s{I}}\!\! \lb \jmp{\b{Q}}  \cdot \o{\b{N}} \rb \, \lb  \avg{\delta \Theta}_{\o{\alpha}} - \delta\o{ \Theta } \rb \, \d A 
\!-\!
\int_{\s{I}}\!\! \f{1}{\o{m}} \, \jmp{\Theta}\, \jmp{\delta \Theta}\, \d A 
\!=\!
\int_{\p\s{B}}\!\! Q^{\text{p}}\, \delta \Theta \, \d A\,.
\end{aligned}
}
\label{eq:FEM-overall}
\end{equation}
Following the interface temperature definition~\eqref{eq:Theta-avg}, the interface test function can be expressed as
\begin{equation}
\begin{aligned}
\delta\o{\Theta}
&=\delta\lp\f{\Theta^{+}\Theta^{-}}{\avg{\Theta}_{1-\o{\alpha}}}\rp
=\f{\lb\Theta^{+}\delta\Theta^{-}+\Theta^{-}\delta\Theta^{+}\rb\avg{\Theta}_{1-\o{\alpha}}}{\lb\avg{\Theta}_{1-\o{\alpha}}\rb^2}-\f{\avg{\delta\Theta}_{1-\o{\alpha}}\Theta^{-}\Theta^{+}}{\lb\avg{\Theta}_{1-\o{\alpha}}\rb^2 } \,,
\end{aligned}
\end{equation}
which can be further simplified to the form
\begin{equation}
\boxed{
\begin{aligned}
\delta\o{\Theta} &= \lb\f{\Theta^{-}}{\avg{\Theta}_{[1-\o{\alpha}]}}-\f{\lb1-\o{\alpha}\rb\Theta^{-}\Theta^{+}}{\lb\avg{\Theta}_{[1-\o{\alpha}]}\rb^2}\rb \delta\Theta^{+} + \lb\f{\Theta^{+}}{\avg{\Theta}_{[1-\o{\alpha}]}}-\f{\o{\alpha}\,\Theta^{-}\Theta^{+}}{\lb\avg{\Theta}_{[1-\o{\alpha}]}\rb^2}\rb
\delta\Theta^{-}\,.
\end{aligned}
}
\label{eq:deltaTheta-interface}
\end{equation}
For $\o{\alpha}=0.5$, Eq.~\eqref{eq:deltaTheta-interface} reduces to
\begin{equation}
\delta\o{\Theta}
=\f{2\lb\Theta^{-}\rb^{2}}{\lb\Theta^{+}+\Theta^{-}\rb^{2}}\,\delta\Theta^{+}
+\f{2\lb\Theta^{+}\rb^{2}}{\lb\Theta^{+}+\Theta^{-}\rb^{2}}\,\delta\Theta^{-}\,.
\label{eq:deltaTheta-interface-half}
\end{equation}

\begin{figure}[b!]
\centering	
\includegraphics[width=0.9\textwidth]{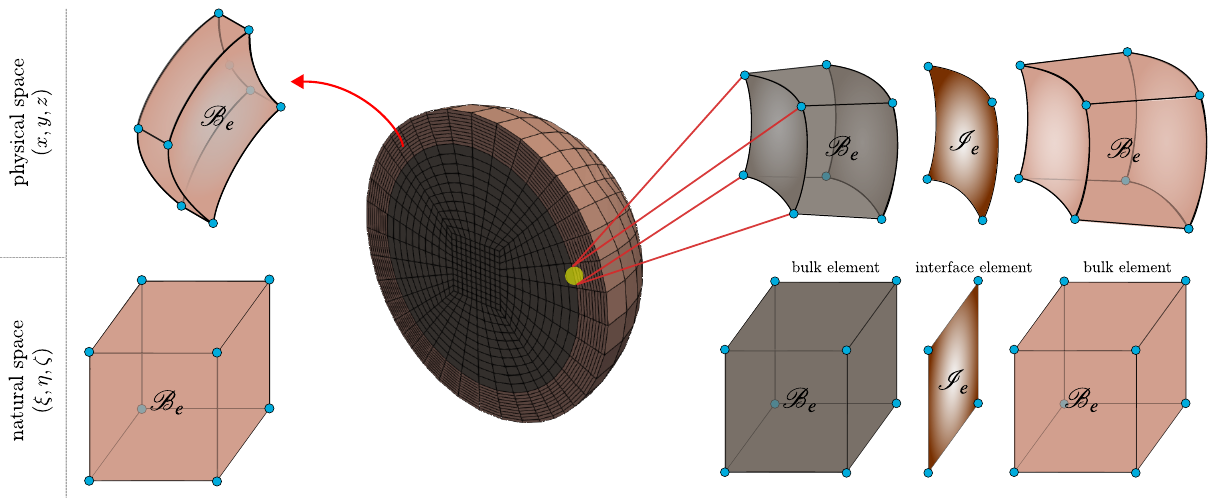}
\caption{
Discretization of bulk and interface elements. 
The interface elements are conformed to the facets of adjacent bulk elements, ensuring that the number of nodes per interface element matches that on each bulk facet. 
This alignment enforces geometrical and interpolation consistency across element boundaries, simplifying implementation without additional degrees of freedom.
}
\label{fig:mesh}
\end{figure}

The next step is discretization which is carried out using the Bubnov--Galerkin scheme.
The interface elements are here taken to be consistent with the bulk elements.
That is, the common face of two bulk elements can be treated as an interface element; for example in 3D, the bulk is discretized with tri-linear elements and the interface with bi-linear elements, while in 2D, the bulk uses bi-linear elements and the interface linear elements.
This choice simplifies the implementation, as the faces of adjacent bulk elements can be directly treated as the two sides of an interface element.
Thus, no extra interpolation or hanging nodes are needed to connect the interface with the bulk.
The bulk and interface domains are discretized into a set of bulk and interface elements as
\begin{equation}
\begin{aligned}
&\s{B} \approx \s{B}^{h} = \assembly{e=1}{\#\text{BE}} \s{B}_{e}\,,
\qquad \qquad 
\p\s{B} \approx \p\s{B}^{h} = \assembly{e=1}{\#\text{BBE}} \p\s{B}_{e}\,,\\
\qquad
&\s{I} \approx \s{I}^{h} = \assembly{e=1}{\#\text{IE}} \s{I}_{e}\,,
\qquad \qquad 
\p\s{I} \approx \p\s{I}^{h} = \assembly{e=1}{\#\text{IBE}} \p\s{I}_{e}\,,
\end{aligned}
\end{equation}
with $\#\text{BE}$ being the number of bulk elements, $\#\text{BBE}$ being the number of bulk boundary elements, $\#\text{IE}$ being the number of interface elements, $\#\text{IBE}$ being the number of interface boundary elements and $\assembly{}{}$ the generic assembly operator.
Utilizing standard interpolations and the isoparametric concept, the geometry, temperature, and test functions associated with bulk and interface elements are approximated as
\begin{equation}
\begin{aligned}
&\b{X}|_{\cc{B}_{e}} \approx \b{X}^{h}(\b{\xi}) =\!\! \sum_{i=1}^{\#\text{NPBE}} \!\! N_{i}(\b{\xi})\, \b{X}_{i}\,,
\quad
&&\o{\b{X}}|_{\cc{I}_{e}} \approx \o{\b{X}}^{h}(\o{\b{\xi}}) =\!\! \sum_{i=1}^{\#\text{NPIE}} \!\! \o{N}_{i}(\o{\b{\xi}})\, \o{\b{X}}_{i}\,,\\
&\Theta|_{\cc{B}_{e}} \approx \Theta^{h}(\b{\xi}) =\!\! \sum_{i=1}^{\#\text{NPBE}} \!\! N_{i}(\b{\xi})\, \Theta_{i}\,,
\quad
&&\o{\Theta}|_{\cc{I}_{e}} \approx \o{\Theta}^{h}(\o{\b{\xi}}) =\!\! \sum_{i=1}^{\#\text{NPIE}} \!\! \o{N}_{i}(\o{\b{\xi}})\, \o{\Theta}_{i}\,,\\
&\delta \Theta|_{\cc{B}_{e}} \approx \delta\Theta^{h}(\b{\xi}) =\!\! \sum_{i=1}^{\#\text{NPBE}} \!\! N_{i}(\b{\xi})\, \delta\Theta_{i}\,,
\quad
&&\delta\o{\Theta}|_{\cc{I}_{e}} \approx \delta\o{\Theta}^{h}(\o{\b{\xi}}) =\!\! \sum_{i=1}^{\#\text{NPIE}} \!\! \o{N}_{i}(\o{\b{\xi}})\, \delta\o{\Theta}_{i}\,,
\end{aligned}
\label{Eq:approx_1}
\end{equation}
with $\#\text{NPBE}$ and $\#\text{NPIE}$ being the number of nodes per bulk element and per interface element, respectively.
The functions $N_{i}$ and $\o{N}_{i}$ are the shape functions of the bulk and interface elements at their local nodes $i$, respectively.
For the bulk elements, $N_{i}$ is defined over the natural coordinate space $\b{\xi} \in [-1,1]^{\text{PD}}$, while for the interface elements, $\o{N}_{i}$ is defined over $\b{\xi} \in [-1,1]^{\text{PD}-1}$, with PD indicating the problem dimension.
Fig.~\ref{fig:mesh} shows the discretization of bulk and interface elements in both two- and three-dimensional settings.
As illustrated, the number of nodes per interface element ($\#\text{NPIE}$) matches the number of nodes on each facet of the adjacent bulk elements.
This ensures conformity between bulk and interface discretizations, avoids the need for special transition elements, and simplifies the numerical implementation.
Following Eqs.~\eqref{Eq:approx_1}, the interpolations for interfacial jump and average read
\begin{equation}
\begin{aligned}
&\jmp{\Theta^{h}} =\!\! \sum_{i=1}^{\#\text{NPIE}} \!\! \o{N}_{i}\,\jmp{\Theta_{i}}  
=\!\!\! \sum_{i=1}^{\#\text{NPIE}} \!\! \o{N}_{i}\,\Theta^{+}_{i} 
\!\!-\!\!\! \sum_{i=1}^{\#\text{NPIE}} \!\! \o{N}_{i}\,\Theta^{-}_{i}\,,
&&\avg{\Theta^{h}}_{\o{\alpha}} =\!\!\!\!\! \sum_{i=1}^{\#\text{NPIE}} \!\! \o{N}_{i}\,\avg{\Theta_{i}}_{\o{\alpha}} 
= \o{\alpha}\!\! \sum_{i=1}^{\#\text{NPIE}} \!\! \o{N}_{i}\,\Theta^{+}_{i}
+ \lb1-\o{\alpha}\rb\!\! \sum_{i=1}^{\#\text{NPIE}} \!\! \o{N}_{i}\,\Theta^{-}_{i} \,,\\
&\jmp{\delta\Theta^{h}} =\!\! \sum_{i=1}^{\#\text{NPIE}} \!\! \o{N}_{i}\,\jmp{\delta\Theta_{i}}  
=\!\!\! \sum_{i=1}^{\#\text{NPIE}} \!\! \o{N}_{i}\,\delta\Theta^{+}_{i} 
\!\!-\!\!\! \sum_{i=1}^{\#\text{NPIE}} \!\! \o{N}_{i}\,\delta\Theta^{-}_{i}\,,
&&\avg{\delta\Theta^{h}}_{\o{\alpha}} =\!\!\!\!\! \sum_{i=1}^{\#\text{NPIE}} \!\! \o{N}_{i}\,\avg{\delta\Theta_{i}}_{\o{\alpha}}  
= \o{\alpha}\!\! \sum_{i=1}^{\#\text{NPIE}} \!\! \o{N}_{i}\,\delta\Theta^{+}_{i}
+ \lb1-\o{\alpha}\rb\!\! \sum_{i=1}^{\#\text{NPIE}} \!\! \o{N}_{i}\,\delta\Theta^{-}_{i} \,,
\end{aligned}
\label{Eq:approx_2}
\end{equation}
where
\begin{equation}
\delta\Theta^{+} = \delta\Theta\big|_{\cc{I}^{+}}\,, \quad \delta\Theta^{-} = \delta\Theta\big|_{\cc{I}^{-}}\,,  \quad \Theta^{+} = \Theta\big|_{\cc{I}^{+}}\,,                         \quad \Theta^{-} = \Theta\big|_{\cc{I}^{-}}\,.
\end{equation}
Using these spatial approximations, the fully-discrete weak form reads
\begin{equation}
\begin{aligned}
&
\assembly{e=1}{\#\text{BE}} \int_{\s{B}_{e}}\!\!\!   
\b{Q} \cdot \lb \sum_{i=1}^{\#\text{NPBE}} \Grad N_{i} \, \delta \Theta_{i}  \rb \,   
\d V  
+
\assembly{e=1}{\#\text{IE}} \int_{\s{I}_{e}}\!\!\!   
\o{\b{Q}} \cdot \lb   \sum_{i=1}^{\#\text{NPIE}} \o{\Grad} \, \o{N}_{i} \, \delta \o{\Theta}_{i} \, \rb 
\d A\\
+&
\assembly{e=1}{\#\text{IE}} \int_{\s{I}_{e}}\!    
\lb \jmp{\b{Q}}  \cdot \o{\b{N}} \rb  
\lb
\o{\alpha}\sum_{i=1}^{\#\text{NPIE}} \!\! \o{N}_{i}\,\delta\Theta^{+}_{i} 
\!\!+ \lb1-\o{\alpha} \rb\sum_{i=1}^{\#\text{NPIE}} \!\! \o{N}_{i}\,\delta\Theta^{-}_{i} - \delta\o{ \Theta}_i
\rb 
 \, \d A  \\
-&\assembly{e=1}{\#\text{IE}} \int_{\s{I}_{e}}\! 
\f{1}{\o{m}} \, \jmp{ \Theta } \lb \, \sum_{i=1}^{\#\text{NPIE}} \!\!\! \o{N}_{i} \delta \Theta^{+}_{i} - \sum_{i=1}^{\#\text{NPIE}} \!\!\! \o{N}_{i} \delta \Theta^{-}_{i}  \, \rb 
\d A  
= 
\assembly{e=1}{\#\text{BBE}} \int_{\p \s{B}_{e}}\!\!\!  
Q^{\text{p}} \lb \sum_{i=1}^{\#\text{NPBBE}} N_{i} \, \delta \Theta_{i}  \rb \,  
\d A  \,,
\end{aligned}
\label{eq:FEM-semi-final}
\end{equation}
which, utilizing the relation~\eqref{eq:deltaTheta-interface}, can be expressed as
\begin{equation}
\hspace{-6pt}
\boxed{
\begin{aligned}
&
\assembly{e=1}{\#\text{BE}} \int_{\s{B}_{e}}\!\!\!   
\b{Q} \cdot \lb \sum_{i=1}^{\#\text{NPBE}} \Grad N_{i} \, \delta \Theta_{i}  \rb \,   
\d V\\  
+&
\assembly{e=1}{\#\text{IE}} \int_{\s{I}_{e}}\!\!\!   
\o{\b{Q}} \cdot \lb  \lb\f{\Theta^{-}_{i}}{\avg{\Theta}_{[1-\o{\alpha}]}}-\f{\lb1-\o{\alpha}\rb\Theta^{-}_{i}\Theta^{+}_{i}}{\lb\avg{\Theta}_{[1-\o{\alpha}]}\rb^2}\rb 
\sum_{i=1}^{\#\text{NPIE}} \o{\Grad} \, \o{N}_{i} \, 
\delta\Theta^{+}_{i}
+
\lb\f{\Theta^{+}_{i}}{\avg{\Theta}_{[1-\o{\alpha}]}}-\f{\o{\alpha}\,\Theta^{-}_{i}\Theta^{+}_{i}}{\lb\avg{\Theta}_{[1-\o{\alpha}]}\rb^2}\rb
\sum_{i=1}^{\#\text{NPIE}} \o{\Grad} \, \o{N}_{i}
\delta\Theta^{-}_{i} \,
\rb 
\d A\\
+&
\assembly{e=1}{\#\text{IE}} \int_{\s{I}_{e}}\!    
\lb \jmp{\b{Q}}  \cdot \o{\b{N}} \rb  
\lb
\lb\o{\alpha} \!-\! \f{\Theta^{-}_{i}}{\avg{\Theta}_{[1-\o{\alpha}]}}\!+\!\f{\lb1-\o{\alpha}\rb\Theta^{-}_{i}\Theta^{+}_{i}}{\lb\avg{\Theta}_{[1\!-\!\o{\alpha}]}\rb^2}\rb
\sum_{i=1}^{\#\text{NPIE}} \!\!\! \o{N}_{i} \delta\Theta^{+}_{i}
\!+\!
\lb[1\!-\!\o{\alpha}]\!-\!\f{\Theta^{+}_{i}}{\avg{\Theta}_{[1-\o{\alpha}]}}\!+\!\f{\o{\alpha}\,\Theta^{-}_{i}\Theta^{+}_{i}}{\lb\avg{\Theta}_{[1-\o{\alpha}]}\rb^2}\rb
\sum_{i=1}^{\#\text{NPIE}} \!\!\! \o{N}_{i}\delta\Theta^{-}_{i}
\rb 
 \, \d A  \\
-&\assembly{e=1}{\#\text{IE}} \int_{\s{I}_{e}}\! 
\f{1}{\o{m}} \, \jmp{ \Theta } \lb \, \sum_{i=1}^{\#\text{NPIE}} \!\!\! \o{N}_{i} \delta \Theta^{+}_{i} - \sum_{i=1}^{\#\text{NPIE}} \!\!\! \o{N}_{i} \delta \Theta^{-}_{i}  \, \rb 
\d A  
= 
\assembly{e=1}{\#\text{BBE}} \int_{\p \s{B}_{e}}\!\!\!  
Q^{\text{p}} \lb \sum_{i=1}^{\#\text{NPBBE}} N_{i} \, \delta \Theta_{i}  \rb \,  
\d A  \,.
\end{aligned}
}
\label{eq:FEM-final}
\end{equation}
Accordingly, the global residual vector is composed of the contributions associated with the bulk temperature field and the two traces of the temperature field across the interface.
Upon assembling the corresponding global residual vector and global temperature field, the resulting nonlinear system of equations is solved iteratively using the Newton--Raphson method.
The computational analysis is carried out using our in-house finite element code applied to the RVE discretized by bi- and tri-linear Lagrange elements as depicted in Fig.~\ref{fig:mesh}.

\section{Homogenization}\label{sec:homogenization}

The aim of homogenization is to derive effective macroscopic quantities that capture the overall response of a heterogeneous medium. 
Central to this is the micro-to-macro transition, in which macroscopic quantities are obtained from their microscopic counterparts by averaging over the {\rve}. 
Denoting by $\b{x}$ an arbitrary point in $\s{B}$, the macroscopic thermal intensity $\M{\b{E}}$ and the macroscopic heat flux $\M{\b{Q}}$ are expressed as volume and interface integrals
\begin{equation}
\M{\b{E}} = \f{1}{\s{V}}\int_\s{B} \b{E}\,\d V + \f{1}{\s{V}}\int_\s{I} \lb \jmp{\Theta} \, \o{\b{N}} \rb  \, \d A\,,
\qquad \qquad
\M{\b{Q}} = \f{1}{\s{V}}\int_\s{B} \b{Q}\,\d V + \f{1}{\s{V}}\int_\s{I} \o{\b{Q}} \, \d A\,,
\label{eq:macro1}
\end{equation}
which can be written in terms of boundary integrals as
\begin{equation}
\M{\b{E}}  = \f{1}{\s{V}}\int_{\p\s{B}} \Theta \, \b{N} \, \d A\,, 
\qquad \qquad
\M{\b{Q}}  = \f{1}{\s{V}}\int_{\p\s{B}} [\b{Q}\cdot \b{N}]\, \b{X} \, \d A\,.
\label{eq:macro2}
\end{equation}
Next we employ the Hill-Mandel condition in order to impose incremental dissipation equivalence between the macro- and micro-scales.
Following the work of Nguyen et al.~\cite{Nguyen2017}, we adopt the generalized Hill--Mandel condition which takes into account the influence of thermal imperfect interfaces as
\begin{equation}
\f{1}{\s{V}}\int_{\s{B}} \b{Q} \cdot \delta\b{E} \,\d V + \f{1}{\s{V}}\int_{\s{I}} \lb[ \o{\b{Q}} \cdot \delta\o{\b{E}} + \o{q}\, \delta\jmp{\Theta} \rb \,\d A - \M{\b{Q}} \cdot \delta\M{\b{E}} \overset{!}{=} 0\,,
\label{Eq:Hill-Mandel-1}
\end{equation}
with $\o{q}=\avg{\b{Q}}_{\lb1-\o{\alpha}\rb}\cdot\o{\b{N}}$ being the normal heat flux across the interface.
Finally, inserting Eq.~\eqref{eq:macro2} into Eq.~\eqref{Eq:Hill-Mandel-1} furnishes the extended Hill--Mandel condition in terms of the boundary integral
\begin{equation}
\int_{\p\s{B}}\lb \de\Theta - \de\M{\b{E}}  \cdot \b{X}  \rb  \cdot  \lb q - \M{\b{Q}} \cdot \b{N} \rb \, \d{A}\,\overset{!}{=}\,0\,,
\label{Eq:Hill-Mandel-2}
\end{equation}
with $q=\b{Q}\cdot \b{N}$ being the bulk normal heat flux.
Among all the boundary conditions that satisfy the Hill--Mandel condition~\eqref{Eq:Hill-Mandel-2}, of practical significance are
\begin{itemize}
    \item Uniform thermal intensity boundary condition: $ \qquad \Theta = \, \M{\b{E}}\cdot \b{X} \quad \text{on}\,\, \p\s{B}\,.$
    \item Constant normal heat flux boundary condition: $ \qquad q =  \M{\b{Q}}\cdot \b{N} \quad \text{on}\,\, \p\s{B}\,.$
    \item Periodic boundary condition: 
    $\qquad \Theta  - \, \M{\b{E}}\cdot \b{x} \quad \text{periodic on}\quad \, \p\s{B}\,, \quad
    q -  \M{\b{Q}}\cdot \b{N}        \quad \text{anti-periodic on}\,\, \p\s{B}\,.$
\end{itemize}
This choice of boundary conditions is suitable for comparisons between the computational and analytical solutions.
See Firooz et al.~\cite{Firooz2019a,Firooz2019} for a comprehensive study on the influence of boundary conditions and the {\rve} types on the behavior of heterogeneous media within the framework of homogenization.

\section{Analytical estimates on the effective conductivity of composites}\label{sec:Analytical} 

This section presents analytical methods to provide estimates for the effective thermal conductivity of composites embedding extended general interfaces.  
The focus lies on the composite assemblage method~\cite{hashin1968assessment}, which is enhanced here to account for extended general interfaces.
In doing so, the composite assemblage approach is here integrated with the Galerkin--Fourier method, whereby the temperature field within each constituent is represented as a Fourier series and the corresponding coefficients are determined by enforcing the boundary and extended general interface conditions in a Galerkin, weighted-residual sense~\cite{BenvenisteMiloh1986c}.
The composite assemblage method is based on explicitly modeling the composite microstructure through idealized, periodic RVEs, such as concentric cylinders or spheres.
Unlike the original variational approach~\cite{Hashin1962a,Hashin1962c,Hashin1963,Hill1963}, which imposes no restriction on the shapes of heterogeneities, the composite assemblage method requires explicit morphological and geometrical details of the microstructure.
It captures phase interactions through exact or semi-analytical solutions to boundary value problems within the RVEs.
Further studies on extensions, connections, and applications of the composite assemblage method in thermal problems can be found in~\cite{Gibiansky1997,Kerner1956a,Rosen1970,Pierard2004,Bohm1994} among others.
Throughout this section, quantities associated with the fiber and the matrix are denoted by superscripts $(1)$ and $(2)$, respectively.

\subsection{Preliminaries}\label{sec:PRE} 

Since fiber- and particle-reinforced composites are used as reference geometries throughout this section, it is useful to first introduce the governing equations in the appropriate coordinate systems.
Specifically, we present the balance and constitutive equations in cylindrical coordinates for fiber-reinforced composites and in spherical coordinates for particle-reinforced composites.
These formulations provide the foundation for all subsequent analytical approaches presented here and allow for the consistent treatment of extended general interface conditions across different micromechanical approaches.

\subsubsection{Cylindrical coordinates}\label{sec:CYL}

\begin{figure}[b!]
	\centering
	{\includegraphics[width=1.0\textwidth]{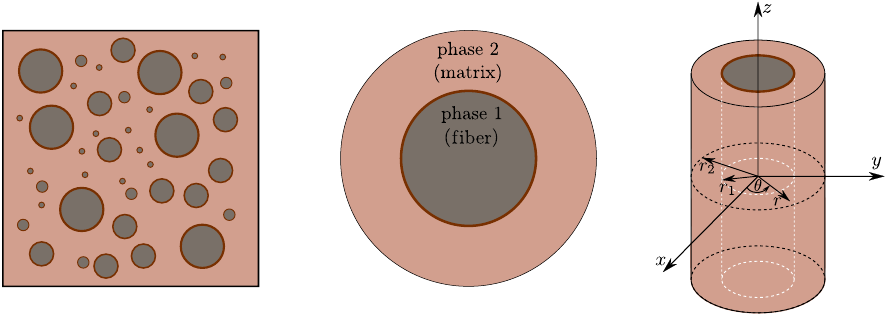}}
	\caption{
	Schematic representation: (left) heterogeneous medium, (center) representative volume element (RVE), and (right) associated coordinate system used to analyze the relations in a transversely isotropic framework.
	}
	\label{fig:CCA-coordinate}
\end{figure}

Fig.~\ref{fig:CCA-coordinate} presents a heterogeneous medium (left) together with its corresponding simplified {\rve} (middle) and the a proper cylindrical coordinate system to analyze such a medium (right). 
The fiber's volume fraction is $f = r_1^2/r_2^2$.
The normal basis vectors in the cylindrical coordinate system is represented as
\begin{equation}
\b{N}_r \equalhat \begin{bmatrix} \cos \theta \\ \sin \theta \\ 0 \end{bmatrix}, \quad
\b{N}_\theta \equalhat \begin{bmatrix} -\sin \theta \\ \cos \theta \\ 0 \end{bmatrix}, \quad
\b{N}_z \equalhat \begin{bmatrix} 0 \\ 0 \\ 1 \end{bmatrix}\,.
\label{eq:basis}
\end{equation}
Thus the heat flux vectors can be expressed as 
\begin{equation}
\begin{aligned}
&\b{Q} = Q_r \b{N}_r +  Q_{\theta} \b{N}_{\theta} + Q_z \b{N}_z, \\
&\o{\b{Q}} = \o{Q}_{\theta} \b{N}_{\theta} + \o{Q}_z \b{N}_z\,.
\end{aligned}
\label{eq:basis2}
\end{equation}
We start by rewriting the heat balance equations given in~\eqref{eq:heat-balance-local} in the cylindrical coordinate system as
\begin{equation}
\begin{aligned}
&\f{\p Q_r}{\p r} + \f{1}{r} \f{\p Q_{\theta}}{\p \theta} + \f{\p Q_z}{\p z} + \f{Q_r}{r} = 0 \, \qquad &&\text{in}\,\,\s{B}\,, \\
&\f{1}{r_1} \f{\p \o{Q}_{\theta}}{\p \theta} + \f{\p \o{Q}_z}{\p z} + \jmp{{Q}_r}= 0\,           \qquad &&\text{on}\,\,\s{I}\,. \\
\end{aligned}
\label{eq:CCAvoight2}
\end{equation}
Next, the constitutive material behavior needs to be defined.
While for the bulk the constitutive material response is standard, for the interface the constitutive material behavior is decomposed into along the interface and across the interface.
That is
\begin{equation}
\boxed{
\begin{aligned}
&\begin{bmatrix}
  Q_r \\
  Q_{\theta} \\
  Q_z \\
\end{bmatrix}
=
\begin{bmatrix}
  K_{r} & 0 & 0 \\
  0 & K_{\theta} & 0 \\
  0 & 0 & K_{z} \\
\end{bmatrix}
\begin{bmatrix}
  E_r \\
  E_{\theta} \\
  E_z \\
\end{bmatrix} 
\, \qquad &&\text{in}\,\,\s{B} && \text{with }\qquad  
E_r = -\f{\p{\Theta}}{\p{r}}\qquad
E_{\theta} = -\f{1}{r}\f{\p{\Theta}}{\p{\theta}}\qquad
E_z = -\f{\p{\Theta}}{\p{Z}}\,,\\
&\begin{bmatrix}
  \o{Q}_{\theta} \\
  \o{Q}_z \\
\end{bmatrix}
=
\begin{bmatrix}
  \o{K}_{\theta} & 0 \\
  0 & \o{K}_{z} \\
\end{bmatrix}
\begin{bmatrix}
  \o{E}_{\theta} \\
  \o{E}_z \\
\end{bmatrix}
\, \qquad &&\text{along}\,\,\s{I} &&\text{with }\qquad  
\o{E}_{\theta} = -\f{1}{r_1}\f{\p{\o{\Theta}}}{\p{\theta}}\qquad
\o{E}_z = -\f{\p{\o{\Theta}}}{\p{z}}\,,\\
&\avg{Q_r}_{[1-\o{\alpha}]} = -\f{1}{\o{m}}\jmp{\Theta} 
\, \qquad &&\text{across}\,\s{I}\,, && \\
\end{aligned}
\label{eq:CCAvoight1}
}
\end{equation}
where $K_{r}$, $K_{\theta}$, $K_{z}$ are the bulk thermal conductivity in the radial, angular and axial directions, respectively, $\o{K}_{\theta}$, $\o{K}_{z}$ are the interface thermal conductivity in the angular and axial directions, respectively and $\o{m}$ is the Kapitza heat flux resistance coefficient. 

\begin{figure}[b!]
	\centering
	{\includegraphics[width=1.0\textwidth]{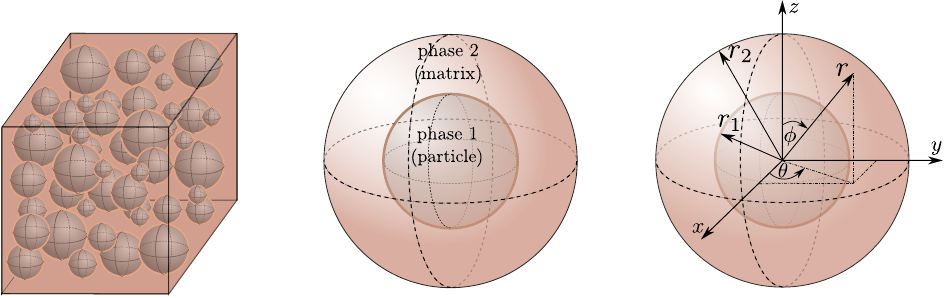}}
	\caption{
	Schematic representation: (left) heterogeneous medium, (center) representative volume element (RVE), and (right) associated coordinate system used to analyze the relations in a spherical coordinate system.
	}
	\label{fig:CSA-coordinate}
\end{figure}

\subsubsection{Spherical coordinate}\label{sec:SPH} 

Fig.~\ref{fig:CSA-coordinate} presents a heterogeneous medium (left) together with its corresponding simplified {\rve} (middle) and the a proper spherical coordinate system to analyze such a medium (right). 
The particle's volume fraction is $f = r_1^3/r_2^3$.
The normal basis vectors in the spherical coordinate system is represented as
\begin{equation}
\b{N}_r \equalhat \begin{bmatrix} \cos \theta \\ \sin \theta \\ 0 \end{bmatrix}, \quad
\b{N}_\theta \equalhat \begin{bmatrix} -\sin \theta \\ \cos \theta \\ 0 \end{bmatrix}, \quad
\b{N}_z \equalhat \begin{bmatrix} 0 \\ 0 \\ 1 \end{bmatrix}\,.
\label{eq:CSAbasis}
\end{equation}
Thus the heat flux vectors can be expressed as
\begin{equation}
\begin{aligned}
&\b{Q} = Q_r \b{N}_r +  Q_{\theta} \b{N}_{\theta} + Q_z \b{N}_z, \\
&\o{\b{Q}} = \o{Q}_{\theta} \b{N}_{\theta} + \o{Q}_z \b{N}_z\,.
\end{aligned}
\label{eq:CSAbasis2}
\end{equation}
Similar to the 2D case, we start by rewriting the bulk and the interface heat balance equations given in~\eqref{eq:heat-balance-local} in the spherical coordinate system as
\begin{equation}
\begin{aligned}
&\f{\p Q_r}{\p r} + \f{1}{r} \f{\p Q_{\theta}}{\p \theta} + \f{1}{r \sin(\theta)}\f{\p Q_{\phi}}{\p \phi} + \f{2Q_r}{r} + \f{Q_{\theta} \, \cos(\theta)}{r \, \sin(\theta)} = 0  \, \qquad &&\text{in}\,\,\s{B}\,,  \\
&\f{1}{r_1} \f{\p \o{Q}_{\theta}}{\p \theta} + \f{1}{r_1 \sin(\theta)} \f{\p \o{Q}_\phi}{\p \phi} + \f{\o{Q}_{\theta} \, \cos(\theta)}{r_1 \, \sin(\theta)} + \jmp{{Q}_r}= 0 \, \qquad &&\text{on}\,\,\s{I}\,.
\end{aligned}
\label{eq:CSAinterfacevoight2}
\end{equation}
Next, the constitutive material behavior needs to be defined.
The bulk constitutive material response is standard and the interface constitutive material behavior is decomposed into along the interface and across the interface.
That is
\begin{equation}
	\hspace{-6pt}
\boxed{
\begin{aligned}
&\begin{bmatrix}
  Q_r \\
  Q_{\theta} \\
  Q_{\phi} \\
\end{bmatrix}
=
\begin{bmatrix}
  K_{r} & 0 & 0 \\
  0 & K_{\theta} & 0 \\
  0 & 0 & K_{\phi} \\
\end{bmatrix}
\begin{bmatrix}
  E_r \\
  E_{\theta} \\
  E_{\phi} \\
\end{bmatrix} 
\, \qquad &&\text{in}\,\,\s{B} &&\text{with }\qquad  
E_r = -\f{\p{\Theta}}{\p{r}}\qquad
E_{\theta} = -\f{1}{r}\f{\p{\Theta}}{\p{\theta}}\qquad
E_{\phi} = -\f{1}{r \sin(\theta)}\f{\p{\Theta}}{\p{\phi}}\,,
\\
&\begin{bmatrix}
  \o{Q}_{\theta} \\
  \o{Q}_{\phi} \\
\end{bmatrix}
=
\begin{bmatrix}
  \o{K}_{\theta} & 0 \\
  0 & \o{K}_{\phi} \\
\end{bmatrix}
\begin{bmatrix}
  \o{E}_{\theta} \\
  \o{E}_{\phi} \\
\end{bmatrix} 
\, \qquad &&\text{along}\,\,\s{I} &&\text{with }\qquad  
\o{E}_{\theta} = -\f{1}{r_1}\f{\p{\o{\Theta}}}{\p{\theta}}\qquad
\o{E}_{\phi} = -\f{1}{r_1 \sin(\theta)}\f{\p{\o{\Theta}}}{\p{\phi}}\,,
\\
&\avg{Q_r}_{[1-\o{\alpha}]} = -\f{1}{\o{m}}\jmp{\Theta}  \, \qquad &&\text{on} \,\,\s{I}\,, &&   
\end{aligned}
}
\label{eq:CSAvoight1}
\end{equation}
where $K_{r}$, $K_{\theta}$, $K_{\phi}$ are the bulk thermal conductivity in the radial, polar and azimuthal directions, respectively, $\o{K}_{\theta}$, $\o{K}_{\phi}$ represent the interface thermal conductivity in the polar and azimuthal directions, respectively and and $\o{m}$ is the Kapitza heat flux resistance coefficient.


\subsection{Composite assemblage method}\label{sec:CAM} 

The composite assemblage method itself can be divided into two categories, depending on the problem of interest:  
(i) the \emph{composite cylinder assemblage} (CCA), which models fiber-reinforced composites in a two-dimensional (2D) axisymmetric setting, and  
(ii) the \emph{composite sphere assemblage} (CSA), which is used for particle-reinforced composites in three dimensions (3D).  
In what follows, we extend both the CCA and the CSA to account for our extended general interface model using the Galerkin--Fourier method, henceforth referred to as the GF-CCA and the GF-CSA, respectively.


\subsubsection{Galerkin--Fourier composite cylinder assemblage}\label{sec:Fiber-Reinforced Composites}

This section aims to develop analytical estimates for the effective conductivity of fiber-reinforced composites embedding extended general interfaces.
In doing so, we present a Galerkin--Fourier composite cylinder assemblage (GF-CCA) approach to account for extended general interfaces allowing for unequal contributions of bulk material to interface properties.
First, we begin by introducing the preliminaries of the RVE problem for fiber-reinforced composites.
This is then followed by extending the GF-CCA approach to account for semi-disspative extended general interfaces.
To do so, we define the average thermal energy in the composite as %
\footnote{The term \emph{energy} is used intentionally to maintain consistency with the mechanical formulation. 
In a purely thermal (steady-state conduction) setting, the corresponding quantity is more naturally interpreted as the dissipation (power), i.e.\ $\mathcal{D} = -U$.}
\begin{equation}
\begin{aligned}
U &= \f{1}{2\s{V}}\int_\s{B} \b{E}\cdot \b{Q} \, \d V + \f{1}{2\s{V}}\int_\s{I} \jmp{-\Theta \,\b{Q}}\cdot \o{\b{N}} \, \d A\,, \\
&= \f{1}{2\s{V}}\int_\s{B} \Div(-\Theta\,\b{Q}) \, \d V + \f{1}{2\s{V}}\int_\s{I} \jmp{-\Theta \,\b{Q}}\cdot \o{\b{N}} \, \d A\,,
\end{aligned}
\label{eq:CCAenergyrve1}
\end{equation}
which, using the extended divergence theorem, updates to
\begin{equation}
\begin{aligned}
U &= -\f{1}{2\s{V}}\int_\s{S}  \Theta \lb\b{Q}\cdot\b{N}\rb \, \,\d A\,.
\end{aligned}
\label{eq:CCAenergyrve2}
\end{equation}
In the cylindrical coordinate system, the volume element is $\d V = r\,\d r\,\d \theta\,\d z$, the surface element at a constant radius $r$ is $\d s_r = r\,\d \theta\,\d z$ and the surface element at a constant height $z$ is $\d s_z=r\,\d r\,\d \theta$. 
Under the prescribed boundary conditions associated with a uniform thermal intensity field, the average thermal energy in the RVE reads
\begin{equation}
\begin{aligned}
U^{\text{RVE}} &= -\f{1}{4\pi {r_2}^2 L}\int_{-L}^{L}\int_{0}^{2\pi} \lb\Theta^{(2)}{Q_r}^{(2)}\rb_{r=r_2}r_2\,\d \theta\,\d z\,,
\end{aligned}
\label{eq:CCAenergyrve3}
\end{equation}
and the average thermal energy in the equivalent homogeneous medium is given as
\begin{equation}
\begin{aligned}
U^{\text{EQ}} &= -\f{1}{4\pi {r_2}^2 L}\int_{-L}^{L}\int_{0}^{2\pi} \lb\Theta^{(\text{eq})}{Q_r}^{(\text{eq})}\rb_{r=r_2}r_2\,\,\d \theta\,\,\d z\,.
\end{aligned}
\label{eq:CCAenergyrve4}
\end{equation}
By equating the energies in~\eqref{eq:CCAenergyrve3} and~\eqref{eq:CCAenergyrve4}, the overall macroscopic thermal conductivity can be obtained. 
Note that the generic expression for the average thermal energy density generally includes additional heat flux components; however, these terms vanish under the uniform thermal intensity field boundary conditions considered here.

In order to calculate the effective conductivity $\M{K}$ in the GF-CCA approach, the {\rve} is subject to a linear temperature field corresponding to a uniform macroscopic temperature gradient prescribed on its boundary as
\begin{equation}
\begin{aligned}
\Theta^{(0)} = \Theta_{\text{ref}} + \beta \, r_2 \, \cos(\theta)\,,
\end{aligned}
\label{eq:ccaN2-1}
\end{equation}
where $\Theta_{\text{ref}}$ is a common reference temperature and $\beta$ denotes the magnitude of the prescribed macroscopic thermal intensity.
Since the interface temperature is defined such that it depends nonlinearly on the bulk temperatures, a purely first-order angular representation is, in general, not sufficient to satisfy the interface conditions.
That is, the nonlinear interface relation may generate higher-order contributions even when the externally prescribed temperature field is first order.
Therefore, the classical first-order composite assemblage approximation needs to be extended to account for higher orders.
Accordingly, we consider a general representation for the evolved temperature field within each constituent, retaining terms up to order $N$, as
\begin{equation}
\begin{aligned}
\Theta^{(1)}
&=
\Theta_0
+
\beta r_1
\sum_{n=1}^{N}
\lb
A_n
\lb\f{r}{r_1}\rb^n
+
B_n
\lb\f{r_1}{r}\rb^n
\rb
\cos(n\theta)\,,
\\
\Theta^{(2)}
&=
\Theta_0
+
\beta r_1
\sum_{n=1}^{N}
\lb
C_n
\lb\f{r}{r_1}\rb^n
+
D_n
\lb\f{r_1}{r}\rb^n
\rb
\cos(n\theta)\,.
\end{aligned}
\label{eq:ccaN2-2}
\end{equation}
Eq.~\eqref{eq:ccaN2-2} constitutes a general representation of the temperature field for an arbitrary truncation order $N$, with the classical CCA form~\cite{Hashin1983} recovered for $N=1$.
Here, we consider its extended version with $N=2$, which yields
\begin{equation}
\begin{aligned}
\Theta^{(1)}
&=
\Theta_0
+
\beta r_1
\lb
A_1\lb\f{r}{r_1}\rb\cos(\theta)
+
B_1\lb\f{r_1}{r}\rb\cos(\theta)
+
A_2\lb\f{r}{r_1}\rb^2\cos(2\theta)
+
B_2\lb\f{r_1}{r}\rb^2\cos(2\theta)
\rb\,,
\\
\Theta^{(2)}
&=
\Theta_0
+
\beta r_1
\lb
C_1\lb\f{r}{r_1}\rb\cos(\theta)
+
D_1\lb\f{r_1}{r}\rb\cos(\theta)
+
C_2\lb\f{r}{r_1}\rb^2\cos(2\theta)
+
D_2\lb\f{r_1}{r}\rb^2\cos(2\theta)
\rb\,.
\end{aligned}
\label{eq:ccaN2-3}
\end{equation}
The constants $A_{1}$, $A_{2}$, $B_{1}$, $B_{2}$, $C_{1}$, $C_{2}$, $D_{1}$ and $D_{2}$ are determined using the boundary and interface conditions;
\begin{itemize}
\item Finite temperature at the center of the fiber
\begin{equation}
\Theta^{(1)} \, \text{finite at } r=0 
\quad \Rightarrow \quad 
B_{1} = 0\,, \qquad \text{and} \qquad B_{2} = 0\,.
\label{eq:cca3}
\end{equation}
\item Prescribed temperature at $r=r_2$
\begin{equation}
\Theta^{(2)}(r_2) = \Theta^{(0)}
\quad \Rightarrow \quad 
D_1=\f{1-C_1}{f}\,, \qquad \text{and} \qquad D_2=-\f{C_2}{f^2}\,.
\label{eq:cca6}
\end{equation}
Accordingly, the coefficients $D_1$ and $D_2$ are expressed in terms of $C_1$ and $C_2$, and the number of independent unknowns reduces to $A_1$, $A_2$, $C_1$, and $C_2$.
\item Equilibrium across the interface at $r=r_1$
\begin{equation}
\avg{Q_{r}{(r_1)}}_{[1-\o{\alpha}]} = -\f{1}{\o{m}}\jmp{\Theta(r_1)} 
\quad \Rightarrow \quad
[1-\o{\alpha}]Q_r^{(2)} + \o{\alpha}Q_r^{(1)} = -\f{1}{\o{m}}[\Theta^{(2)}(r_1) - \Theta^{(1)}(r_1)]\,.
\label{eq:cca4}
\end{equation}
Substitution of the temperature fields~\eqref{eq:ccaN2-3} into Eq.~\eqref{eq:cca4} gives
\begin{equation}
\cc{R}_{n_1} \cos(\Theta) + \cc{R}_{n_2} \cos(2\Theta) = 0\,.
\label{eq:cca4-2}
\end{equation}
Since $\cos(\theta)$ and $\cos(2\theta)$ are orthogonal over the interval
$[0,2\pi]$, we have
\begin{equation}
\int_{0}^{2\pi}
\cos(\theta)\cos(2\theta)
\,\mathrm{d}\theta
=
0\,.
\label{eq:cca4-orthogonality}
\end{equation}
Therefore, the two angular contributions in Eq.~\eqref{eq:cca4-2} are independent. 
In order for Eq.~\eqref{eq:cca4-2} to hold for all $\theta$, the coefficient of each angular mode must vanish separately, such that $\cc{R}_{n_1}=0$ and $\cc{R}_{n_2}=0$, which leads to
\begin{equation}
\begin{aligned}
\cc{R}_{n_1}
&=
-[1-\o{\alpha}]K^{(2)}\lb C_1-\f{1-C_1}{f}\rb
-\o{\alpha}K^{(1)}A_1
+
\f{r_1}{\o{m}}
\lb
C_1+\f{1-C_1}{f}-A_1
\rb
=0\,,
\\
\cc{R}_{n_2}
&=
-2[1-\o{\alpha}]K^{(2)}\lb C_2+\f{C_2}{f^2}\rb
-2\o{\alpha}K^{(1)}A_2
+
\f{r_1}{\o{m}}
\lb
C_2-\f{C_2}{f^2}-A_2
\rb
=0\,.
\end{aligned}
\label{eq:ccaN2-normal-system}
\end{equation}

\item Equilibrium along the interface at $r=r_1$
\begin{equation}
[\o{\Div}\,\o{Q}]_r + \jmp{Q_r} = 0
\quad \Rightarrow \quad
\f{1}{r_1}\f{\p \o{Q}_{\theta}}{\p \theta}
+
\f{\p \o{Q}_z}{\p z}
+
\jmp{Q_r(r_1)}
=
0\,,
\label{eq:cca5}
\end{equation}
which yields
\begin{equation}
Q_r^{(2)}
-
Q_r^{(1)}
-
\f{\o{K}}{r_1^2}
\f{\p^2 \o{\Theta}}{\p \theta^2}
=
0\,.
\label{eq:ccaN2-Ft}
\end{equation}
The interface temperature $\o{\Theta}$ is obtained by substituting the temperature fields~\eqref{eq:ccaN2-3} into the interface temperature definition~\eqref{eq:Theta-avg}, which reads
\begin{equation}
\o{\Theta}(\theta)
=
\f{
N_0
+
N_1\cos(\theta)
+
N_2\cos(2\theta)
+
N_3\cos(3\theta)
+
N_4\cos(4\theta)
}{
M_0
+
M_1\cos(\theta)
+
M_2\cos(2\theta)
}\,,
\label{eq:ccaN2-harmonic-expanded}
\end{equation}
where the coefficients $N_0,\ldots,N_4$ and $M_0,M_1,M_2$ are given in~\ref{sec:app_1}.
It is observed that the bulk approximations contain only the first two angular modes, whereas the interface temperature contains additional modes generated through weighted harmonic averaging.
As a result, the tangential equilibrium condition~\eqref{eq:ccaN2-Ft} cannot, in general, be satisfied pointwise for every value of $\theta$ using only the four unknown coefficients $A_1$, $A_2$, $C_1$, and $C_2$. 
Therefore, we enforce the tangential equilibrium condition in the space spanned by the two angular modes retained in the bulk approximations.
To this end, Eq.~\eqref{eq:ccaN2-Ft} is written in a form of residual projected onto $\cos(\theta)$ and $\cos(2\theta)$ such that $\cc{R}_{t_1}=0$ and $\cc{R}_{t_2}=0$. 
The corresponding residual equations are defined as
\begin{equation}
\begin{aligned}
\cc{R}_{t_1}
&=
\f{1}{\pi}
\int_{0}^{2\pi}
\lb Q_r^{(2)}
-
Q_r^{(1)}
-
\f{\o{K}}{r_1^2}
\f{\p^2 \o{\Theta}}{\p \theta^2} \rb
\cos(\theta)
\,\d\theta
=
0\,,\\
\cc{R}_{t_2}
&=
\f{1}{\pi}
\int_{0}^{2\pi}
\lb Q_r^{(2)}
-
Q_r^{(1)}
-
\f{\o{K}}{r_1^2}
\f{\p^2 \o{\Theta}}{\p \theta^2} \rb
\cos(2\theta)
\,\d\theta
=
0\,.
\end{aligned}
\label{eq:ccaN2-Rt2}
\end{equation}
Using the orthogonality of the trigonometric functions, these equations may equivalently be written as
\begin{equation}
\begin{aligned}
\mathcal{R}_{t}^{(1)}
&=
\beta
\lb
K^{(1)}A_1
-
K^{(2)}\lb C_1-\f{1-C_1}{f}\rb
\rb
+
\f{\o{K}}{\pi r_1^2}
\int_{0}^{2\pi}
\o{\Theta}(\theta)
\cos(\theta)
\,\d\theta
=0\,,
\\
\mathcal{R}_{t}^{(2)}
&=
2\beta
\lb
K^{(1)}A_2
-
K^{(2)}\lb C_2+\f{C_2}{f^2}\rb
\rb
+
\f{4\o{K}}{\pi r_1^2}
\int_{0}^{2\pi}
\o{\Theta}(\theta)
\cos(2\theta)
\,\d\theta
=0\,.
\end{aligned}
\label{eq:ccaN2-tangential-system2}
\end{equation}
The above projections extract the first- and second-order angular components of the tangential residual. 
Owing to the orthogonality of the Fourier modes, the higher-order contributions do not directly contribute to the corresponding projected residual. 
\end{itemize}

The final system can thus be written compactly as
\begin{equation}
\b{R}(\b{x})
=
\lb
\mathcal{R}_{n}^{(1)},
\mathcal{R}_{n}^{(2)},
\mathcal{R}_{t}^{(1)},
\mathcal{R}_{t}^{(2)}
\rb^{\text{T}}
=
\b{0}\,,
\end{equation}
which constitutes a nonlinear system of equations with the unknowns $\b{x}=\lb A_1,\,A_2,\,C_1,\,C_2\rb^{\text{T}}$, solved iteratively using the Newton--Raphson method.
With the unknowns thus determined, the average thermal energies in the {\rve}~\eqref{eq:CCAenergyrve3} and the equivalent homogeneous medium~\eqref{eq:CCAenergyrve4} under prescribed temperature field read
\begin{equation}
\begin{aligned}
&U^{\text{RVE}} = \f{1}{2}\,\beta^{2}K^{(2)}\lb 2C_{1} - 1 \rb\,,\\
&U^{\text{EQ}}  = \f{1}{2}\,\beta^{2}\,\M{K}\,,
\end{aligned}
\label{eq:csaenergy1}
\end{equation}
where $\M{K}$ is the overall effective conductivity.
Imposing $U^{\textrm{(RVE)}}=U^{\textrm{(eq)}}$ results in the overall conductivity $\M{K}$ as
\begin{equation}
\M{K}
=
K^{(2)}
\lb
2C_1-1
\rb\,.
\label{eq:ccaN2-keff2}
\end{equation}


\subsubsection{Galerkin--Fourier composite sphere assemblage}\label{sec:Particle Reinforced Composites}

This section aims to develop analytical estimates for the effective conductivity of particle-reinforced composites embedding extended general interfaces. 
In doing so, we present a Galerkin--Fourier composite sphere assemblage (GF-CSA) approach to account for extended general interfaces allowing for unequal contributions of bulk material to interface properties.

We define the average thermal energy in the composite as 
\begin{equation}
\begin{aligned}
U &= \f{1}{2\s{V}}\int_\s{B} \b{E}\cdot \b{Q} \, \,\d V + \f{1}{2\s{V}}\int_\s{I} \jmp{-\Theta \b{Q}}\cdot \o{\b{N}} \, \,\d A\,, \\
&= \f{1}{2\s{V}}\int_\s{B} \Div(-\Theta\b{Q}) \, \,\d V + \f{1}{2\s{V}}\int_\s{I} \jmp{-\Theta \b{Q}}\cdot \o{\b{N}} \, \,\d A\,,
\end{aligned}
\label{eq:CSAenergyrve1}
\end{equation}
which, using the extended divergence theorem, updates to
\begin{equation}
\begin{aligned}
U &= -\f{1}{2\s{V}}\int_\s{S}  \Theta \lb \b{Q}\cdot\b{N} \rb \, \,\d A\,.
\end{aligned}
\label{eq:CSAenergyrve2}
\end{equation}
In a spherical coordinate system, the volume element is $\,\d V=r^2\, \sin(\theta) \,\d r\,\d \theta\,\,\d \phi $ and the surface element at constant radius $r$ is $\,\d s_r=r^2\, \sin(\theta) \,\d \theta\,\d \phi $. 
Under the prescribed boundary conditions associated with uniform thermal intensity field, the average thermal energy in the RVE reads 
\begin{equation}
\begin{aligned}
U^{\text{RVE}} &= -\f{3}{8\pi r_2}\int_{0}^{2 \pi}\int_{0}^{\pi} \lb\Theta^{(2)}{Q_r}^{(2)}\rb_{r=r_2} \sin(\theta) \,\d \theta\,\d \phi\,,
\end{aligned}
\label{eq:CSAenergyrve3}
\end{equation}
and the average thermal energy in the equivalent homogeneous medium is given as
\begin{equation}
\begin{aligned}
U^{\text{EQ}} &= -\f{3}{8\pi r_2}\int_{0}^{2 \pi}\int_{0}^{\pi} \lb \Theta^{(\text{eq})}{Q_r}^{(\text{eq})}\rb_{r=r_2} sin(\theta) \,\d \theta\,\d \phi \,.
\end{aligned}
\label{eq:CSAenergyrve4}
\end{equation}
By equating the energies in~\eqref{eq:CSAenergyrve3} and~\eqref{eq:CSAenergyrve4}, the overall macroscopic thermal conductivity can be obtained. 
Note that the generic expression for the average thermal energy density generally includes additional heat flux components; however, they vanish under the uniform thermal intensity field boundary conditions considered here.

In order to calculate the effective conductivity $\M{K}$ in the extended Fourier--Galerkin CSA approach, the {\rve} is subject to a linear temperature field corresponding to a uniform macroscopic temperature gradient prescribed on its boundary as
\begin{equation}
\Theta^{(0)}
=
\Theta_{\text{ref}}
+
\beta r_2\cos(\theta)\,,
\label{eq:CSA-N2-1}
\end{equation}
where $\Theta_{\text{ref}}$ is a common reference temperature and $\beta$ denotes the magnitude of the prescribed macroscopic thermal intensity.
Similar to the GF-CCA approach, we consider a general representation for the evolved temperature field within each constituent, retaining terms up to order $N$, as
\begin{equation}
\begin{aligned}
\Theta^{(1)}
&=
\Theta_{\text{ref}}
+
\beta r_1
\sum_{n=1}^{N}
\lb
A_n
\lb\f{r}{r_1}\rb^n
+
B_n
\lb\f{r_1}{r}\rb^{n+1}
\rb
P_n(\cos\theta)\,,
\\
\Theta^{(2)}
&=
\Theta_{\text{ref}}
+
\beta r_1
\sum_{n=1}^{N}
\lb
C_n
\lb\f{r}{r_1}\rb^n
+
D_n
\lb\f{r_1}{r}\rb^{n+1}
\rb
P_n(\cos\theta)\,,
\end{aligned}
\label{eq:CSA-N2-2}
\end{equation}
where $P_n$ denotes the Legendre polynomial of order $n$.
Eq.~\eqref{eq:CSA-N2-2} constitutes a general representation of the temperature field for an arbitrary truncation order $N$, with the classical CSA form~\cite{hashin1968assessment} recovered for $N = 1$. 
Here, we consider its extended version with $N = 2$, which yields
\begin{equation}
\begin{aligned}
\Theta^{(1)}
&=
\Theta_{\text{ref}}
+
\beta r_1
\lb
A_1\f{r}{r_1}P_1(\cos\theta)
+
B_1\lb\f{r_1}{r}\rb^2P_1(\cos\theta)
+
A_2\lb\f{r}{r_1}\rb^2P_2(\cos\theta)
+
B_2\lb\f{r_1}{r}\rb^3P_2(\cos\theta)
\rb\,,
\\
\Theta^{(2)}
&=
\Theta_{\text{ref}}
+
\beta r_1
\lb
C_1\f{r}{r_1}P_1(\cos\theta)
+
D_1\lb\f{r_1}{r}\rb^2P_1(\cos\theta)
+
C_2\lb\f{r}{r_1}\rb^2P_2(\cos\theta)
+
D_2\lb\f{r_1}{r}\rb^3P_2(\cos\theta)
\rb\,,
\end{aligned}
\label{eq:CSA-N2-3}
\end{equation}
with the Legendre polynomials being
\begin{equation}
P_1(\cos\theta)=\cos(\theta)\,,
\qquad
P_2(\cos\theta)=\f{1}{2}\lb3\cos^2(\theta)-1\rb\,.
\label{eq:CSA-Legendre}
\end{equation}
The constants $A_1$, $A_2$, $B_1$, $B_2$, $C_1$, $C_2$, $D_1$ and $D_2$ in Eq.~\eqref{eq:CSA-N2-3} are determined using the boundary and interface conditions;
\begin{itemize}

\item Finite temperature at the center of the particle
\begin{equation}
\Theta^{(1)}\;\text{finite at }r=0
\quad\Rightarrow\quad
B_1=0\,,
\qquad\text{and}\qquad
B_2=0\,.
\label{eq:CSA3-FG}
\end{equation}

\item Prescribed temperature at $r=r_2$
\begin{equation}
\Theta^{(2)}(r_2)=\Theta^{(0)}
\quad\Rightarrow\quad
D_1=\f{1-C_1}{f}\,,
\qquad\text{and}\qquad
D_2=-\f{C_2}{f^{5/3}}\,.
\label{eq:CSA6-FG}
\end{equation}
Accordingly, the coefficients $D_1$ and $D_2$ are expressed in terms of $C_1$ and $C_2$, and the number of independent unknowns reduces to $A_1$, $A_2$, $C_1$ and $C_2$.

\item Equilibrium across the interface at $r=r_1$
\begin{equation}
\avg{Q_r(r_1)}_{[1-\o{\alpha}]}
=
-\f{1}{\o{m}}\jmp{\Theta(r_1)}
\quad\Rightarrow\quad
[1-\o{\alpha}]Q_r^{(2)}
+
\o{\alpha}Q_r^{(1)}
=
-\f{1}{\o{m}}
\lb
\Theta^{(2)}(r_1)
-
\Theta^{(1)}(r_1)
\rb\,.
\label{eq:CSA4-FG}
\end{equation}
At the interface, the radial heat fluxes are
\begin{equation}
\begin{aligned}
Q_r^{(1)}
&=
-K^{(1)}\beta
\Big[
A_1P_1(\cos\theta)
+
2A_2P_2(\cos\theta)
\Big]\,,
\\
Q_r^{(2)}
&=
-K^{(2)}\beta
\lb
\lb C_1-\f{2[1-C_1]}{f}\rb P_1(\cos\theta)
+
\lb2C_2+\f{3C_2}{f^{5/3}}\rb P_2(\cos\theta)
\rb\,.
\end{aligned}
\label{eq:CSA-interface-flux-FG}
\end{equation}
Substitution of the temperature fields~\eqref{eq:CSA-N2-3} into Eq.~\eqref{eq:CSA4-FG} gives
\begin{equation}
\cc{R}_{n_1}P_1(\cos\theta)
+
\cc{R}_{n_2}P_2(\cos\theta)
=
0\,.
\label{eq:CSA4-2-FG}
\end{equation}
The Legendre polynomials are orthogonal over the spherical surface such that
\begin{equation}
\int_{0}^{\pi}
P_1(\cos\theta)
P_2(\cos\theta)
\sin(\theta)
\,\d\theta
=
0\,.
\label{eq:CSA4-orthogonality-FG}
\end{equation}
Therefore, the two angular contributions in Eq.~\eqref{eq:CSA4-2-FG} are independent.
In order for Eq.~\eqref{eq:CSA4-2-FG} to hold for all $\theta$, the coefficient of each angular mode must vanish separately, such that $\cc{R}_{n_1}=0$ and $\cc{R}_{n_2}=0$, which leads to
\begin{equation}
\begin{aligned}
\cc{R}_{n_1}
&=
-[1-\o{\alpha}]K^{(2)}
\lb
C_1-\f{2(1-C_1)}{f}
\rb
-\o{\alpha}K^{(1)}A_1
+
\f{r_1}{\o{m}}
\lb[
C_1+\f{1-C_1}{f}-A_1
\rb
=
0\,,
\\
\cc{R}_{n_2}
&=
-[1-\o{\alpha}]K^{(2)}
\lb
2C_2+\f{3C_2}{f^{5/3}}
\rb
-2\o{\alpha}K^{(1)}A_2
+
\f{r_1}{\o{m}}
\lb[
C_2-\f{C_2}{f^{5/3}}-A_2
\rb]
=
0\,.
\end{aligned}
\label{eq:CSA-N2-normal-system}
\end{equation}

\item Equilibrium along the interface at $r=r_1$
\begin{equation}
[\o{\Div}\,\o{\b{Q}}]_r+\jmp{Q_r}=0
\quad\Rightarrow\quad
\f{1}{r_1}\f{\p\o{Q}_{\theta}}{\p\theta}
+
\f{1}{r_1\sin(\theta)}
\f{\p\o{Q}_{\phi}}{\p\phi}
+
\f{\o{Q}_{\theta}\cos(\theta)}
{r_1\sin(\theta)}
+
\jmp{Q_r(r_1)}
=
0\,,
\label{eq:CSA5-FG}
\end{equation}
which yields
\begin{equation}
Q_r^{(2)}
-
Q_r^{(1)}
-
\f{\o{K}}{r_1^2}
\lb
\f{\p^2\o{\Theta}}{\p\theta^2}
+
\cot(\theta)
\f{\p\o{\Theta}}{\p\theta}
\rb
=
0\,.
\label{eq:CSA-N2-Ft}
\end{equation}
The interface temperature $\o{\Theta}$ is obtained by substituting the temperature fields~\eqref{eq:CSA-N2-3} into the weighted harmonic interface temperature definition~\eqref{eq:avg}, which reads
\begin{equation}
\o{\Theta}(\theta)
=
\f{
N_0
+
N_1P_1(\cos\theta)
+
N_2P_2(\cos\theta)
+
N_3P_3(\cos\theta)
+
N_4P_4(\cos\theta)
}{
M_0
+
M_1P_1(\cos\theta)
+
M_2P_2(\cos\theta)
}\,,
\label{eq:CSA-N2-harmonic-expanded}
\end{equation}
where the coefficients $N_0,\ldots,N_4$ and $M_0,M_1,M_2$ are given in~\ref{sec:app_2}.
Similarly, it is observed that the bulk approximations contain only the first two Legendre modes, whereas the interface temperature contains additional modes generated through weighted harmonic averaging.
As a result, the tangential equilibrium condition~\eqref{eq:CSA-N2-Ft} cannot, in general, be satisfied pointwise for every value of $\theta$ using only the four unknown coefficients $A_1$, $A_2$, $C_1$ and $C_2$.
Therefore, we enforce the tangential equilibrium condition in the space spanned by the two Legendre modes retained in the bulk approximation.
To this end, Eq.~\eqref{eq:CSA-N2-Ft} is written in a form of a residual projected onto $P_1(\cos\theta)$ and $P_2(\cos\theta)$ such that $\cc{R}_{t_1}=0$ and $\cc{R}_{t_2}=0$.
The corresponding residual equations are defined as
\begin{equation}
\begin{aligned}
\cc{R}_{t_1}
&=
\f{3}{2}
\int_{0}^{\pi}
\lb
Q_r^{(2)}
-
Q_r^{(1)}
-
\f{\o{K}}{r_1^2}
\lb
\f{\p^2\o{\Theta}}{\p\theta^2}
+
\cot(\theta)
\f{\p\o{\Theta}}{\p\theta}
\rb
\rb
P_1(\cos\theta)
\sin(\theta)
\,\d\theta
=
0\,,
\\
\cc{R}_{t_2}
&=
\f{5}{2}
\int_{0}^{\pi}
\lb
Q_r^{(2)}
-
Q_r^{(1)}
-
\f{\o{K}}{r_1^2}
\lb
\f{\p^2\o{\Theta}}{\p\theta^2}
+
\cot(\theta)
\f{\p\o{\Theta}}{\p\theta}
\rb
\rb
P_2(\cos\theta)
\sin(\theta)
\,\d\theta
=
0\,.
\end{aligned}
\label{eq:CSA-N2-Rt}
\end{equation}
Using the orthogonality of the trigonometric functions and the Legendre differential equation\footnote{$\f{1}{\sin(\theta)} \f{\p}{\p\theta} \lb \sin(\theta) \f{\p P_i(\cos\theta)}{\p\theta} \rb = -i(i+1)P_i(\cos\theta)$}, the tangential residual may equivalently be written as
\begin{equation}
\begin{aligned}
\cc{R}_{t_1}
&=
\beta
\lb
K^{(1)}A_1
-
K^{(2)}\lb C_1-\f{2\lb1-C_1\rb}{f}\rb
\rb
+
\f{3\o{K}}{r_1^2}
\int_{0}^{\pi}
\o{\Theta}(\theta)
P_1(\cos\theta)
\sin(\theta)
\,\d\theta
=
0\,,
\\
\cc{R}_{t_2}
&=
\beta
\lb
2K^{(1)}A_2
-
K^{(2)}\lb2C_2+\f{3C_2}{f^{5/3}}\rb
\rb
+
\f{15\o{K}}{r_1^2}
\int_{0}^{\pi}
\o{\Theta}(\theta)
P_2(\cos\theta)
\sin(\theta)
\,\d\theta
=
0\,.
\end{aligned}
\label{eq:CSA-N2-tangential-system2}
\end{equation}
The above projections extract the first- and second-order Legendre components of the tangential residual.
Owing to the orthogonality of the Legendre modes over the spherical surface, the higher-order contributions do not directly contribute to the corresponding projected residual.
\end{itemize}

The final system can thus be written compactly as
\begin{equation}
\b{R}(\b{x})
=
\lb
\cc{R}_{n_1},
\cc{R}_{n_2},
\cc{R}_{t_1},
\cc{R}_{t_2}
\rb^{\text{T}}
=
\b{0}\,,
\label{eq:CSA-N2-system}
\end{equation}
which constitutes a nonlinear system of equations with the unknowns $\b{x}=\lb A_1,\,A_2,\,C_1,\,C_2\rb^{\text{T}}$, solved iteratively using the Newton--Raphson method.
With the unknowns thus determined, the average thermal energies in the {\rve}~\eqref{eq:CSAenergyrve3} and the equivalent homogeneous medium~\eqref{eq:CSAenergyrve4} under prescribed temperature field read
\begin{equation}
\begin{aligned}
&U^{\text{RVE}}
=
\f{1}{2}\beta^2K^{(2)}
\lb
3C_1-2
\rb\,,
\\
&U^{\text{EQ}}
=
\f{1}{2}\beta^2\M{K}\,,
\end{aligned}
\label{eq:CSAenergy1-FG}
\end{equation}
where $\M{K}$ is the overall effective conductivity.
Imposing $U^{\textrm{(RVE)}}=U^{\textrm{(eq)}}$ results in the overall conductivity $\M{K}$ as
\begin{equation}
\M{K}
=
K^{(2)}
\lb
3C_1-2
\rb
\,.
\label{eq:CSA-N2-keff}
\end{equation}


\section{Numerical examples}\label{sec:Numerical examples}

In this section, we present a comprehensive set of numerical examples to demonstrate the capabilities of the proposed extended general interface model. 
First, we compare the response of different interface models, namely the low conducting (LC) interface, the high conducting (HC) interface, and the general interface model. 
We show that the extended general interface model recovers these models as special cases through a suitable choice of the weighting factor and the interface parameters, and that it additionally provides a continuous transition between them. 
Next, we study the influence of the {\rve} size on the overall conductivity and compare the analytical estimates with the corresponding FEM results. 
Afterwards, we analyze the influence of the interface properties along and across the interface, as well as the inclusion-to-matrix conductivity ratio, on the overall conductivity of fiber- and particle-reinforced composites. 
Finally, we investigate the effectiveness of the proposed interface model in capturing the behavior of composites with graded interphases. 
All FEM results are obtained using our in-house FEM code.


\subsection{Unit definitions}

\begin{figure}[t!]
\centering	
\includegraphics[width=0.9\textwidth]{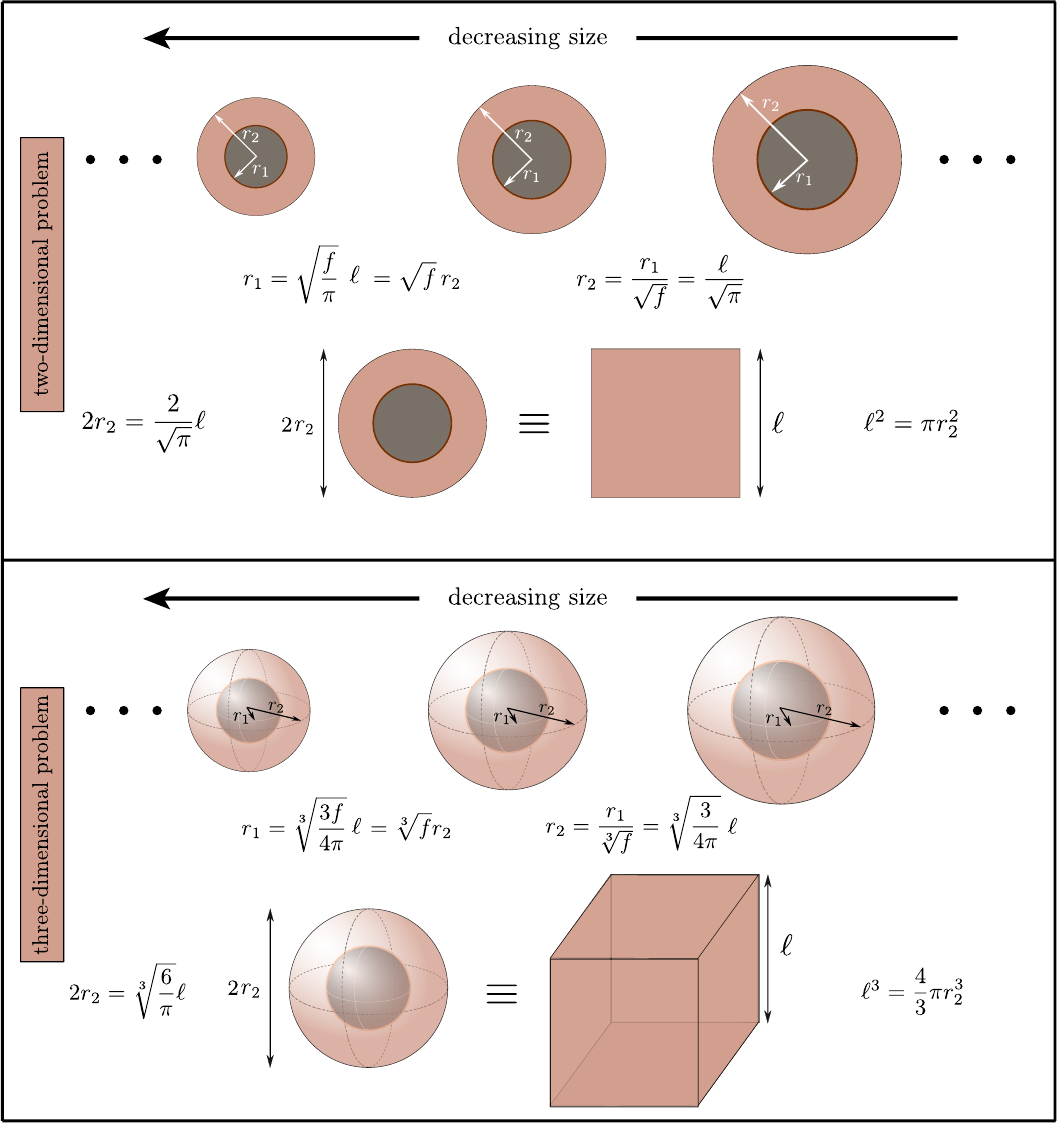}
\caption{
Representative volume element (RVE) illustrating the definition of ``size'': 
(top) two-dimensional case for fiber-reinforced composites, where $\ell2 = \pi r_2^2$, and (bottom) three-dimensional case for particle-reinforced composites, where $\ell^3 = \tfrac{4}{3}\pi r_2^3$.}
\label{fig:size}
\end{figure}

Before proceeding, it is useful to clarify the notion of ``size'' denoted as $\ell$. 
In this context, ``size'' refers to the side length of a square and represents the physical dimensions of the {\rve}, as illustrated in Fig.~\ref{fig:size}. For a given $\ell$ and volume fraction $f$, the radii of the fiber/particle ($r_1$) and the matrix ($r_2$) are determined accordingly. 
The matrix radius is obtained from the area (in 2D) or volume (in 3D) equivalence, i.e., $\ell^2 = \pi r_2^2$ in two dimensions and $\ell^3 = \tfrac{4}{3}\pi r_2^3$ in three dimensions. 
Consequently, both radii decrease as the overall size is reduced.
Assuming that $\ell$ is expressed in mm, the bulk thermal conductivities $K^{(1)}$ and $K^{(2)}$ are measured in $\mathrm{W}/(\mathrm{mm}\mathrm{K})$, 
while the interface conductivity $\o{K}$ is in $\mathrm{W}/\mathrm{K}$. 
The orthogonal interface resistance $\o{m}$ has units $\mathrm{K}\,\mathrm{mm}^2/\mathrm{W}$. 
The weighting factor $\o{\alpha}$ determines how much inclusion and matrix contribute to the interface related quantities. 
In the limiting cases, $\o{\alpha}=0$ corresponds to a case where the interface fully take its associated quantities from the fiber while $\o{\alpha}=1$ corresponds to a case where the matrix fully determines the interface related quantities.  
The conductivity ratio, denoted as $K^{(1)}/K^{(2)}$, is the ratio of the inclusion to the matrix conductivities.


\subsection{Comparison of interface models}

\begin{figure}[b!]
	\centering	
	\includegraphics[width=0.81\textwidth]{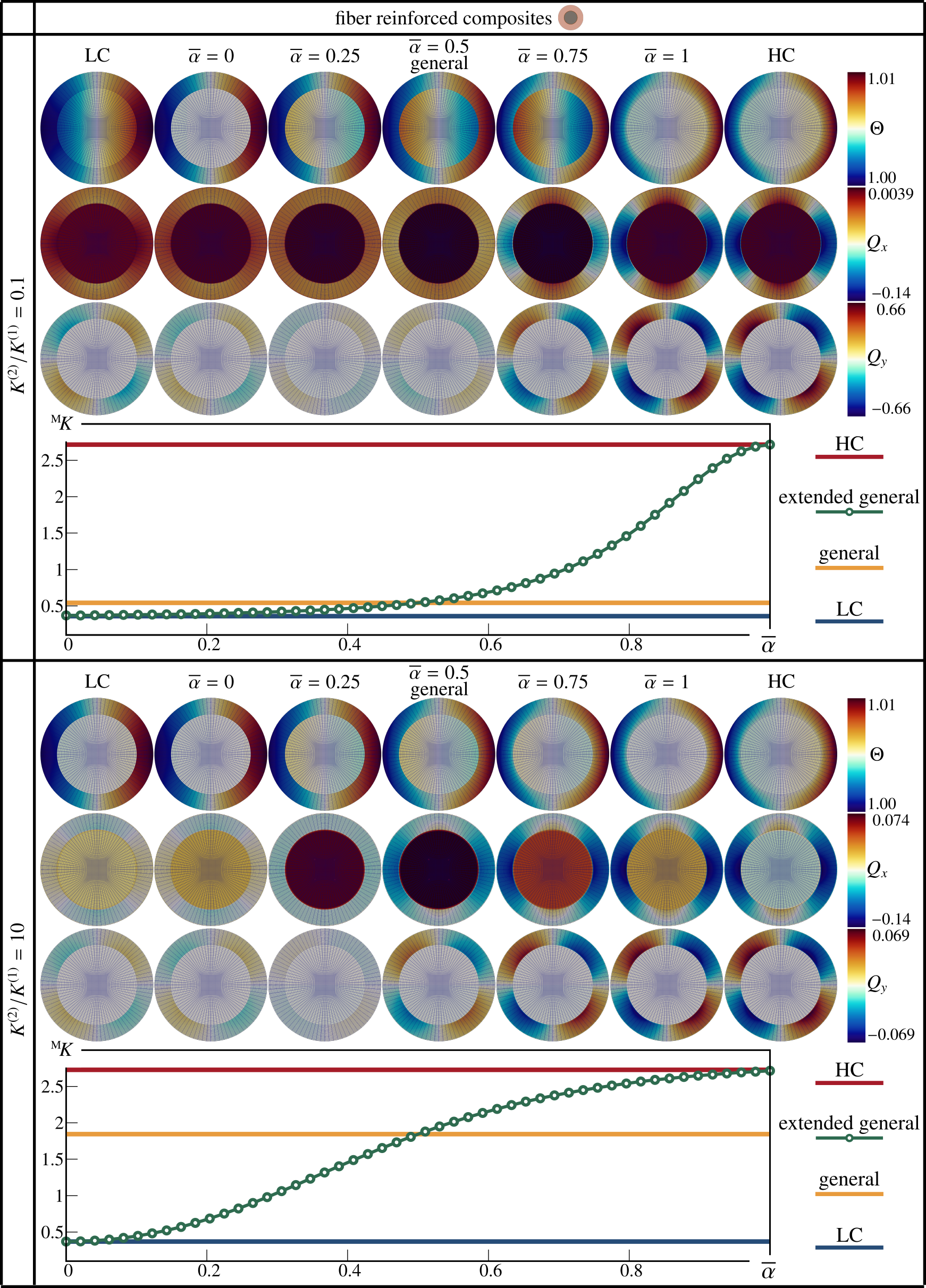}
	\caption{Comparison of interface models for a fiber-reinforced composite with inclusion volume fraction $f=50\%$, {\rve} size $\ell=0.1$, and interface parameters $\o{K}=1.0$ and $\o{m}=1.0$. The top and bottom blocks correspond to the conductivity ratios $K^{(1)}/K^{(2)}=0.1$ and $K^{(1)}/K^{(2)}=10$, respectively. In each block, the first, second, and third rows show the distributions of the temperature $\Theta$ and the heat fluxes $Q_x$ and $Q_y$, respectively, obtained from FEM. The columns show, from left to right, the LC interface ($\o{K}=0$), the extended general interface model with $\o{\alpha}=0,\,0.25,\,0.5,\,0.75,\,1$, and the HC interface ($\o{m}\rightarrow 0$). The plot below the specimens shows the overall conductivity versus the weighting factor $\o{\alpha}$ for the HC (solid red), LC (solid blue), general (orange), and extended general (green with markers) interface models.}
\label{fig:FEM_ANA_2D}
\end{figure}

\begin{figure}[b!]
	\centering	
	\includegraphics[width=0.81\textwidth]{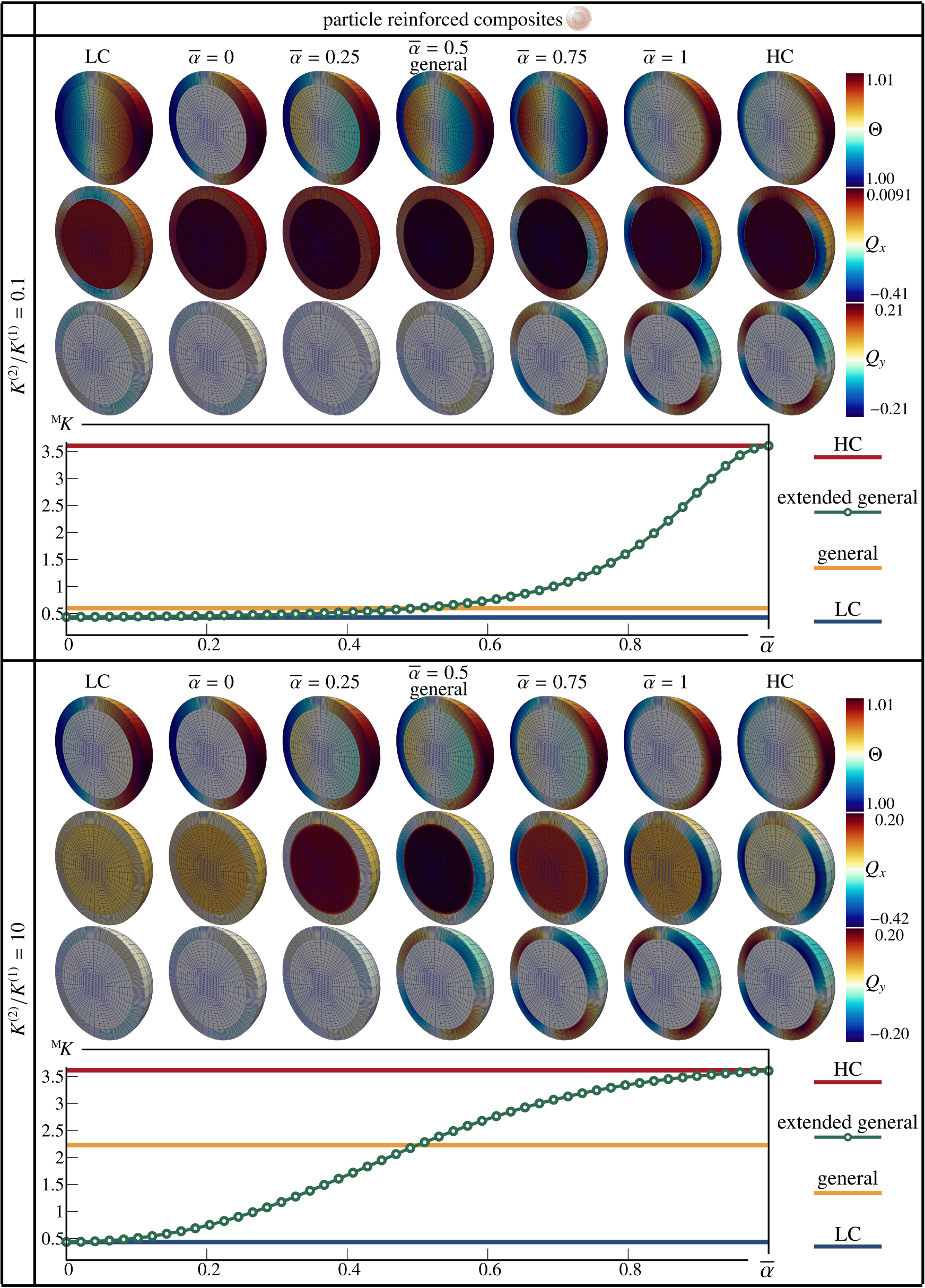}
	\caption{Comparison of interface models for a particle-reinforced composite with inclusion volume fraction $f=50\%$, {\rve} size $\ell=0.1$, and interface parameters $\o{K}=1.0$ and $\o{m}=1.0$. The top and bottom blocks correspond to the conductivity ratios $K^{(1)}/K^{(2)}=0.1$ and $K^{(1)}/K^{(2)}=10$, respectively. In each block, the first, second, and third rows show the distributions of the temperature $\Theta$ and the heat fluxes $Q_x$ and $Q_y$, respectively, obtained from FEM. The columns show, from left to right, the LC interface ($\o{K}=0$), the extended general interface model with $\o{\alpha}=0,\,0.25,\,0.5,\,0.75,\,1$, and the HC interface ($\o{m}\rightarrow 0$). The plot below the specimens shows the overall conductivity versus the weighting factor $\o{\alpha}$ for the HC (solid red), LC (solid blue), general (orange), and extended general (green with markers) interface models.}
\label{fig:FEM_ANA_3D}
\end{figure}

In this example, we compare the response of different interface models and show how the extended general interface model recovers them.
Figures~\ref{fig:FEM_ANA_2D} and~\ref{fig:FEM_ANA_3D} show the results for fiber- and particle-reinforced composites, respectively.
In each figure, two cases are considered: a conductivity ratio of $K^{(1)}/K^{(2)}=0.1$ shown in the top block and $K^{(1)}/K^{(2)}=10$ shown in the bottom block.
In each case, a linear Dirichlet temperature field is prescribed on the boundary of the {\rve} according to $\Theta(r_2,\theta)=\Theta_0+\beta r_2\cos\theta$.
The inclusion volume fraction is $f=50\%$, the {\rve} size is $\ell=0.1$, and the interface parameters are set to $\o{K}=1.0$ and $\o{m}=1.0$.
All results shown in these figures are FEM outputs.
In each block, the first row shows the temperature distribution, the second row shows the heat flux $Q_x$, and the third row shows the heat flux $Q_y$.
Each row contains seven specimens.
The leftmost specimen corresponds to the LC interface, which is recovered from our model by setting $\o{K}=0$.
The next five specimens show the response of the extended general interface model for the weighting factors $\o{\alpha}=0,\,0.25,\,0.5,\,0.75,\,1$, where $\o{\alpha}=0.5$ recovers the classical general interface model.
The rightmost specimen shows the response of the HC interface, which is obtained from our model by letting $\o{m}\rightarrow 0$.
Below the specimens, the overall conductivity is plotted as a function of the weighting factor $\o{\alpha}$ for four interface types.
The solid red line corresponds to the HC interface, the solid blue line to the LC interface, the orange line to the general interface, and the green line with markers to the extended general interface model.

Looking at the temperature distributions, we observe that the temperature jump between the inclusion and the matrix tends to vanish as $\o{\alpha}$ increases.
Thus, although $\o{m}$ does not tend to zero, the weighting factor smooths out the temperature jump.
The opposite holds for the heat fluxes.
As $\o{\alpha}$ increases, the difference between the heat flux on the left and right sides of the inclusion and the heat flux in the adjacent matrix tends to increase.
In the overall conductivity plot, it is observed that LC, HC, and general interface models remain insensitive to the weighting factor, yielding constant curves. 
By contrast, the extended general interface model exhibits strong dependence on the weighting factor, spanning the entire range between the LC and HC bounds. 
Moreover the LC interface provides a lower bound and the HC interface provides an upper bound for the overall conductivity.
The overall response of the extended general interface model coincides with that of the LC interface at $\o{\alpha}=0$, with that of the HC interface at $\o{\alpha}=1$, and with that of the general interface model at $\o{\alpha}=0.5$.
This trend is observed for all cases considered.
Another interesting observation is that, although the temperature and heat flux distributions of the LC interface differ from those of the extended general interface model at $\o{\alpha}=0$, and likewise those of the HC interface differ from those at $\o{\alpha}=1$, the overall responses are the same.

\subsection{Overall conductivity; analytical estimates vs. FEM}

\begin{figure}[b!]
	\centering	
	\includegraphics[width=1.0\textwidth]{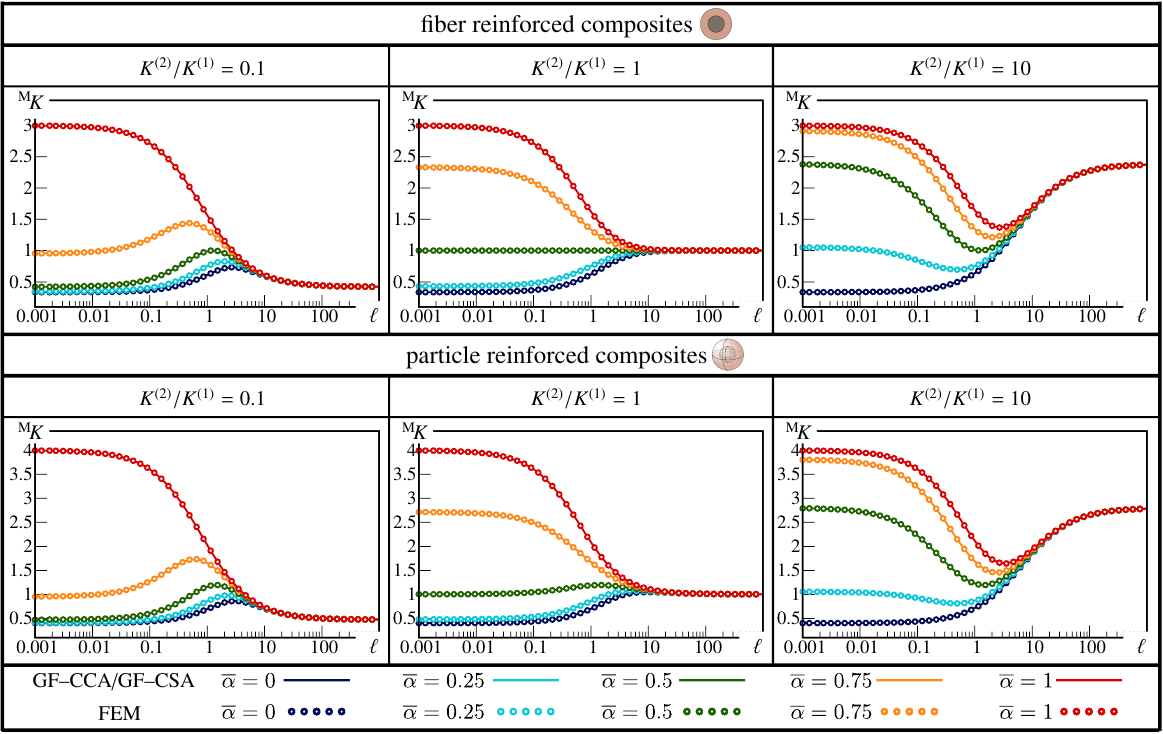}
	\caption{Influence of the {\rve} size $\ell$ on the overall conductivity for fiber-reinforced (top block) and particle-reinforced (bottom block) composites with inclusion volume fraction $f=50\%$, matrix conductivity $K^{(2)}=1$, and interface parameters $\o{K}=1.0$ and $\o{m}=1.0$. Each block contains three panels corresponding to the conductivity ratios $K^{(1)}/K^{(2)}=0.1$, $1$, and $10$. In each panel, the lines show the analytical estimates obtained from the GF-CCA and GF-CSA methods and the points show the FEM results for the weighting factors $\o{\alpha}=0,\,0.25,\,0.5,\,0.75,\,1$.}
\label{fig:SiZe}
\end{figure}

In this example, we investigate how the overall conductivity varies with the {\rve} size.
Moreover, we compare the overall response obtained from the analytical estimates with that obtained from FEM.
Figure~\ref{fig:SiZe} shows the overall conductivity versus the {\rve} size $\ell$.
The top block corresponds to fiber-reinforced composites and the bottom block corresponds to particle-reinforced composites.
Each block contains three panels associated with the conductivity ratios $K^{(1)}/K^{(2)}=0.1$, $1$, and $10$.
In each panel, five lines are plotted, corresponding to the weighting factors $\o{\alpha}=0,\,0.25,\,0.5,\,0.75,\,1$.
The inclusion volume fraction is $f=50\%$, the matrix conductivity is fixed as $K^{(2)}=1$, and the interface parameters are set to $\o{K}=1.0$ and $\o{m}=1.0$.
In each plot, the lines represent the analytical estimates obtained from the GF-CCA and GF-CSA methods, while the points represent the FEM results.
A remarkable agreement between the two approaches is observed.
The analytical estimates coincide with the FEM results for all cases, confirming the accuracy of the proposed GF-CCA and GF-CSA methods.
This agreement holds for all weighting factors, conductivity ratios, and {\rve} sizes considered, for both fiber- and particle-reinforced composites.
It is observed that at large {\rve} sizes, all lines coincide.
This is expected, since the interface effects diminish as the {\rve} size increases, and the responses for all weighting factors converge to a common value.
On the other hand, as the size decreases, the interface effects become dominant and the responses deviate from each other, highlighting the influence of the interface.
For all cases, the response for $\o{\alpha}=0$ gives the lowest overall conductivity, and the overall conductivity increases as $\o{\alpha}$ increases until $\o{\alpha}=1$, which gives the highest.
This is consistent with the previous example, where the LC and HC responses provide the lower and upper bounds, respectively.
Another interesting observation is that, for all cases, the responses for $\o{\alpha}=0$ and $\o{\alpha}=0.25$ yield a smaller-less-conductive response, i.e., the overall conductivity falls below the large-size limit.
In contrast, the responses for $\o{\alpha}=0.75$ and $\o{\alpha}=1$ yield a smaller-more-conductive response, i.e., the overall conductivity lies above the large-size limit.
For $\o{\alpha}=0.5$, the responses at small and large sizes are almost the same.
Moreover, the value to which all the responses converge at large sizes increases with the conductivity ratio.
This is also consistent with our expectation, since only the bulk effects are dominant at such sizes, and with the matrix conductivity fixed, a higher inclusion conductivity leads to a higher overall conductivity.

Since the trends and observations are almost identical for fiber- and particle-reinforced composites, we present the remaining numerical examples only for fiber-reinforced composites for conciseness and to save space.


\subsection{The role of interface parameters}

\begin{figure}[b!]
	\centering	
	\includegraphics[scale=0.73]{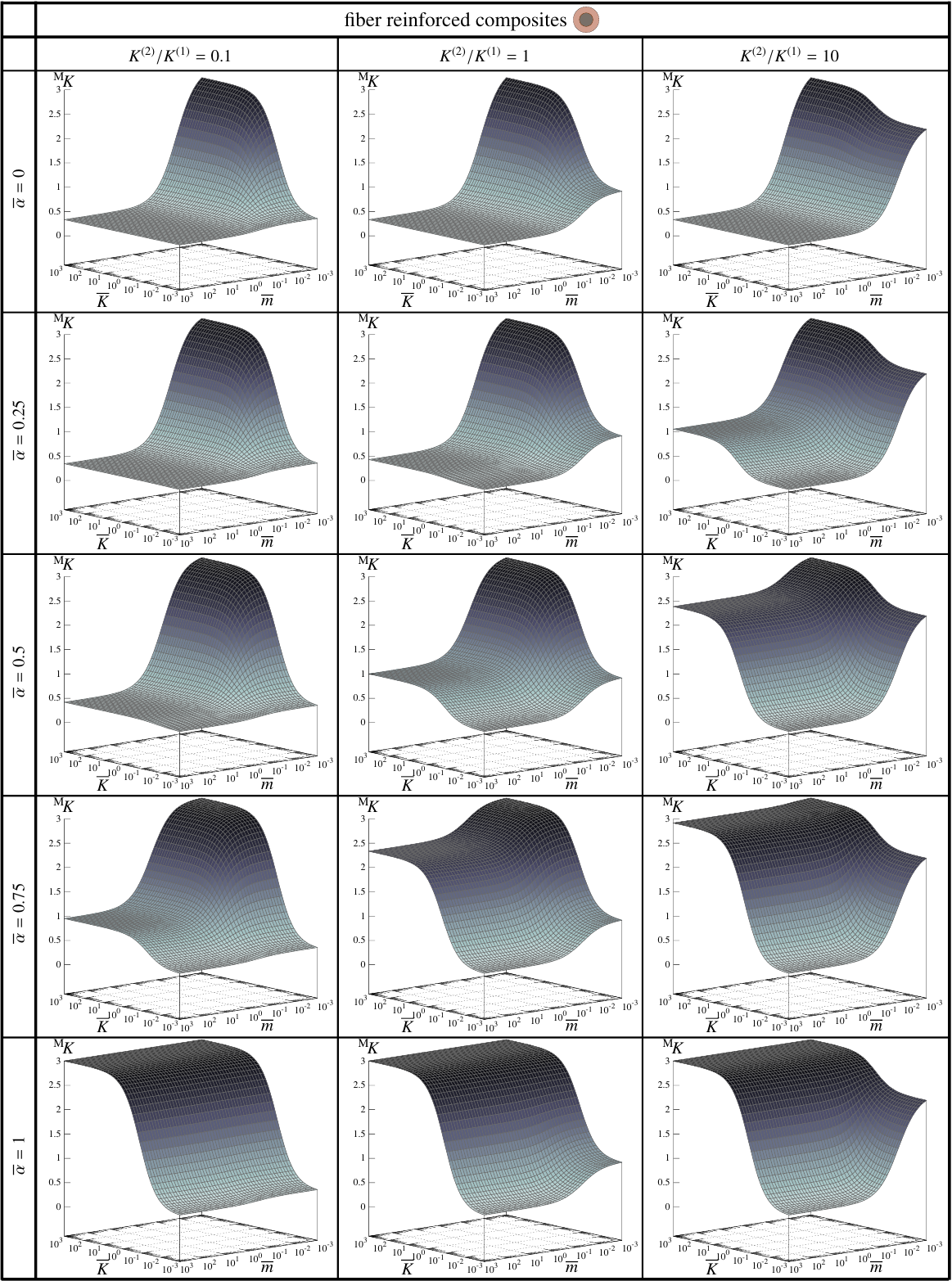}
	\caption{Influence of the interface conductivity $\o{K}$ and the Kapitza resistance $\o{m}$ on the overall conductivity $\M{K}$ of fiber-reinforced composites with inclusion volume fraction $f=50\%$, {\rve} size $\ell=0.1$, and matrix conductivity $K^{(2)}=1$. 
	The columns correspond to the conductivity ratios $K^{(1)}/K^{(2)}=0.1$, $1$, and $10$, and the rows correspond to the weighting factors $\o{\alpha}=0,\,0.25,\,0.5,\,0.75,\,1$.}
\label{fig:Kt_Kn_2D}
\end{figure}

In this example, we examine the influence of the interface conductivity $\o{K}$ and the Kapitza heat flux resistance $\o{m}$ on the effective thermal conductivity of fiber-reinforced composites.
Figure~\ref{fig:Kt_Kn_2D} plots the overall conductivity $\M{K}$ versus the interface conductivity $\o{K}$ and the Kapitza heat flux resistance $\o{m}$.
Each column corresponds to one of the conductivity ratios $K^{(1)}/K^{(2)}=0.1$, $1$, and $10$, and each row corresponds to one of the weighting factors $\o{\alpha}=0,\,0.25,\,0.5,\,0.75,\,1$.
The inclusion volume fraction is $f=50\%$, the {\rve} size is $\ell=0.1$, and the matrix conductivity is $K^{(2)}=1$.

For $\o{\alpha}=0$ and large $\o{m}$, the overall conductivity is almost independent of $\o{K}$.
In contrast, for small $\o{m}$, the overall conductivity depends on $\o{K}$, and larger values of $\o{K}$ lead to larger overall conductivities.
This dependence diminishes as the conductivity ratio increases.
Moreover, for a given $\o{K}$, the overall conductivity increases as $\o{m}$ decreases, because a smaller resistance eases the heat transfer across the interface.
A similar behavior is observed for $\o{\alpha}=0.25$ with the conductivity ratios $0.1$ and $1$.
For $\o{\alpha}=0.25$ and a conductivity ratio of $10$, however, $\M{K}$ varies with $\o{K}$ even at large $\o{m}$, and increasing $\o{K}$ leads to higher overall conductivities.
For $\o{\alpha}=0.5$, the behavior at a conductivity ratio of $0.1$ remains similar to that at $\o{\alpha}=0$.
At the conductivity ratios $1$ and $10$, however, the overall conductivity depends on both $\o{K}$ and $\o{m}$, and large overall conductivities are obtained by increasing $\o{K}$ and decreasing $\o{m}$.
For $\o{\alpha}=0.75$ and a conductivity ratio of $0.1$, the overall conductivity depends on $\o{K}$ for large $\o{m}$.
This dependence intensifies as the conductivity ratio increases.
For a conductivity ratio of $10$ and large $\o{K}$, the overall response is almost independent of $\o{m}$, whereas for small $\o{K}$, decreasing $\o{m}$ yields a larger overall conductivity.
Finally, for $\o{\alpha}=1$ and large $\o{K}$, all cases show almost no dependence on $\o{m}$.
For small $\o{K}$, on the other hand, decreasing $\o{m}$ leads to a larger overall conductivity.
It is important to emphasize the two limiting cases.
For $\o{\alpha}=0$, the overall conductivity does not vary with $\o{K}$ at large $\o{m}$, while for $\o{\alpha}=1$, it does not vary with $\o{m}$ at large $\o{K}$.
In between, the weighting factor gradually shifts the response from the former behavior to the latter.


\subsection{The role of inclusion to matrix conductivity ratio}

\begin{figure}[b!]
	\centering	
	\includegraphics[width=1.0\textwidth]{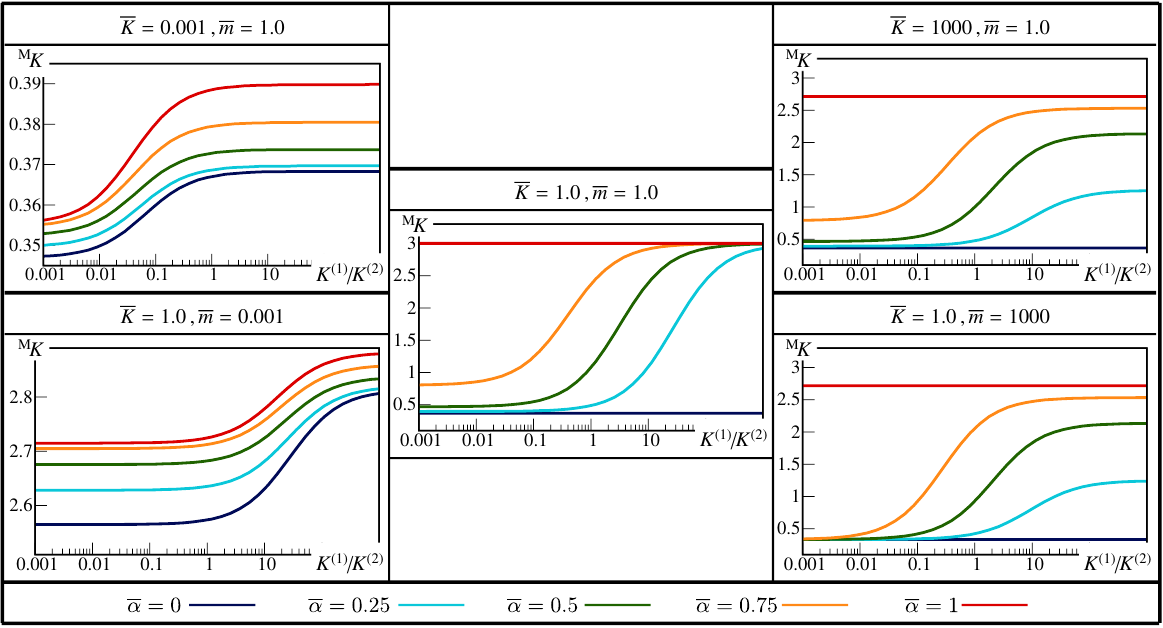}
	\caption{Influence of the inclusion-to-matrix conductivity ratio $K^{(1)}/K^{(2)}$ on the overall conductivity $\M{K}$ of fiber-reinforced composites with inclusion volume fraction $f=50\%$ and {\rve} size $\ell=0.1$. The panels correspond to the interface parameters $\o{K}=0.001,\,\o{m}=1$; $\o{K}=1,\,\o{m}=1$; $\o{K}=1000,\,\o{m}=1$; $\o{K}=1,\,\o{m}=0.001$; and $\o{K}=1,\,\o{m}=1000$. In each panel, the lines show the responses for the weighting factors $\o{\alpha}=0,\,0.25,\,0.5,\,0.75,\,1$.}
\label{fig:SR}
\end{figure}

In this example, we examine the role of the inclusion-to-matrix conductivity ratio $K^{(1)}/K^{(2)}$ in the overall conductivity of fiber-reinforced composites.
Figure~\ref{fig:SR} shows the overall conductivity $\M{K}$ versus the conductivity ratio $K^{(1)}/K^{(2)}$.
The inclusion volume fraction is $f=50\%$ and the {\rve} size is $\ell=0.1$, while $K^{(1)}$ and $K^{(2)}$ are varied to span the range of conductivity ratios.
The figure contains five panels corresponding to five sets of interface parameters: $\o{K}=0.001,\,\o{m}=1$; $\o{K}=1,\,\o{m}=1$; $\o{K}=1000,\,\o{m}=1$; $\o{K}=1,\,\o{m}=0.001$; and $\o{K}=1,\,\o{m}=1000$.
With this parameter, we keep one interface parameter equal to $1$ and assign a very small and a very large value to the other, and we additionally consider the case in which both parameters are equal to $1$.
In each panel, the lines correspond to the weighting factors $\o{\alpha}=0,\,0.25,\,0.5,\,0.75,\,1$.
For all cases, $\o{\alpha}=0$ gives the lowest overall conductivity, and the overall conductivity increases with $\o{\alpha}$ until $\o{\alpha}=1$, which gives the highest overall conductivity.
For $\o{K}=0.001,\,\o{m}=1$ and $\o{K}=1,\,\o{m}=0.001$, the responses for all weighting factors change with the conductivity ratio, and increasing $K^{(1)}/K^{(2)}$ leads to a higher overall conductivity.
For the other three cases, the responses for $\o{\alpha}=0$ and $\o{\alpha}=1$ remain insensitive to the conductivity ratio and provide constant lower and upper bounds, respectively.
The responses for $\o{\alpha}=0.25,\,0.5$, and $0.75$ start close to the response for $\o{\alpha}=0$ at small conductivity ratios and tend towards the response for $\o{\alpha}=1$ at large conductivity ratios.

\subsection{Interface vs interphase}\label{sec:interface_vs_interphase}

\begin{figure}[b!]
	\centering	
	\includegraphics[width=0.84\textwidth]{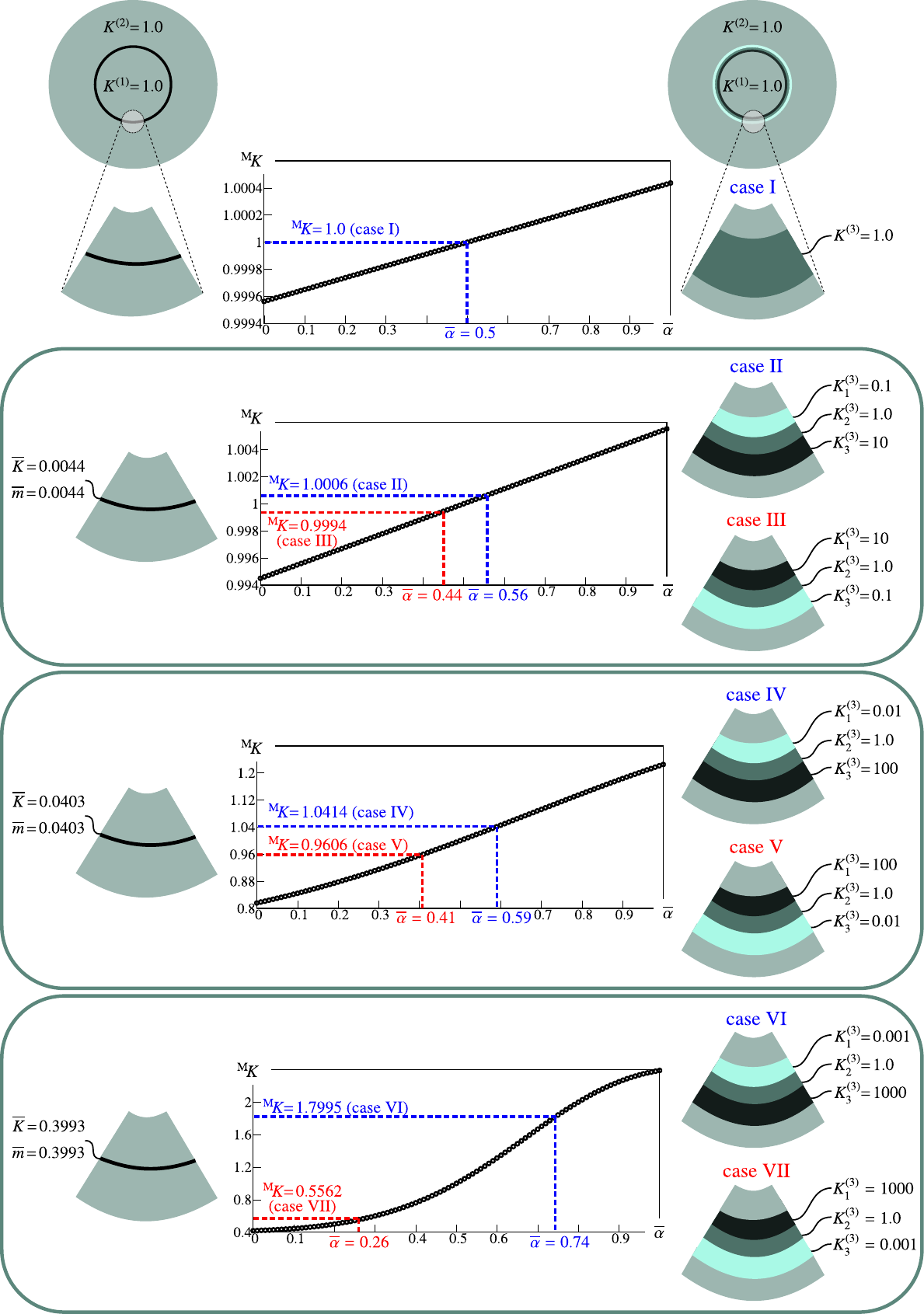}
	\caption{Comparison of the extended general interface model with graded interphase models for fiber-reinforced composites with inclusion volume fraction $f=50\%$ and {\rve} size $\ell=0.1$. 
	The interphase consists of three layers, each with a thickness of $0.01$ times the inclusion radius. 
	For all cases, the overall conductivity $\M{K}$ is plotted versus the weighting factor $\o{\alpha}$ for the iven interface parameters $\o{K}$ and $\o{m}$.}
\label{fig:interphase_interface}
\end{figure}

In this final example, we compare the effective conductivity of a composite medium with a finite interphase to that of a composite medium modeled using the extended general interface. In all cases, the inclusion and matrix conductivities are taken as $K^{(1)}=K^{(2)}=1.0$, the inclusion volume fraction is $f=50\%$, and the {\rve} size is $\ell=0.1$. A three-layer interphase, denoted by the superscript $(3)$, is introduced between the inclusion and the matrix, with each layer having a thickness of $0.01R$, where $R$ denotes the inclusion radius. The corresponding interface model is constructed by replacing the finite interphase with a zero-thickness interface. The interface parameters are determined from the classical series--parallel relations
\begin{equation}
\o{K}=\sum_{i=1}^{3} K_i^{(3)} t_i,
\qquad
\o{m}=\sum_{i=1}^{3} \frac{t_i}{K_i^{(3)}},
\label{eq:interphase_interface_relations}
\end{equation}
where $K_i^{(3)}$ and $t_i$ denote the conductivity and thickness of the $i$th interphase layer, respectively. For all seven cases, the overall conductivity of the composite containing the finite interphase is first obtained using FEM. The overall conductivity of the corresponding interface model is then evaluated as a function of the weighting factor $\o{\alpha}$ using the interface parameters given by Eq.~\eqref{eq:interphase_interface_relations}. The value of $\o{\alpha}$ is identified as the value for which the interface model reproduces the same overall conductivity as the finite interphase model. The resulting comparisons are presented in Fig.~\ref{fig:interphase_interface}.
Seven cases are considered. In Case I, all three interphase layers are identical, with $K_1^{(3)}=K_2^{(3)}=K_3^{(3)}=1.0$.
In this case, the corresponding interface model recovers the same overall conductivity with $\o{\alpha}=0.5$, which is consistent with equal contributions from the inclusion and matrix sides. 
The remaining six cases are arranged in three pairs. 
In each pair, one case places the more conductive interphase layers closer to the inclusion, whereas the other reverses the order.
For Cases II and III, the interphase conductivities are $[\,0.1,\;1,\;10\,]$ and $[\,10,\;1,\;0.1\,]$ listed from the inclusion side to the matrix side.
For this block, the overall conductivity is recovered at $\o{\alpha}=0.55$ for case II and at $\o{\alpha}=0.45$ for case III. 
For Cases IV and V, the interphase conductivities are $[\,0.01,\;1,\;100\,]$ and $[\,100,\;1,\;0.01\,]$.
For this block, the overall conductivity is recovered at $\o{\alpha}=0.59$ for case IV and at $\o{\alpha}=0.41$ for case V. 
Finally, for Cases VI and VII, the interphase conductivities are $[\,0.0001,\;1,\;1000\,]$ and $[\,1000,\;1,\;0.001\,]$ for which the corresponding interface model yields the identical conductivities at $\o{\alpha}=0.74$ and $\o{\alpha}=0.26$, respectively.
Note that in all pairs of cases in each box, the two $\o{\alpha}$ values add up to exactly 1, which is expected since the superposition of each pair must lead to $\o{\alpha}=0.5$.

The results show that the classical choice $\o{\alpha}=0.5$ is appropriate for a uniform interphase, as expected. For graded interphases, however, the matching value of $\o{\alpha}$ generally departs from $0.5$, indicating that equal contributions from the two sides are no longer sufficient, highlighting the necessity of the proposed model.
In particular, the first two graded blocks clearly show that when the more conductive interphase layers are located closer to the matrix side, the identified value satisfies $\o{\alpha}>0.5$, whereas when the more conductive layers are located closer to the inclusion side, the identified value satisfies $\o{\alpha}<0.5$, that was anticipated in Fig.~\ref{fig:Mot}.
This confirms that the weighting factor $\o{\alpha}$ represents the underlying mechanism for incorporating the asymmetry of graded interphases into the equivalent interface description.
Hence, unlike the classical general interface model, the extended general interface model can represent not only uniform interphases, but also complex interphases through an appropriate choice of $\o{\alpha}$.


\section{Conclusion}\label{sec:conclusion}

We developed a novel extended general interface model to account for imperfect bonding in heterogeneous materials.
The central feature of the model is the weighting coefficient $\o{\alpha}$, which allows for unequal contributions from the matrix and the inclusion to the interface fields.
Through thermodynamics considerations, we revisited the governing equations of heterogeneous continua embedding extended general interfaces.
The extended general interface parameters recovers all the previously established models such as the lowly conducting (LC), the highly conducting (HC), and the classical general interface models.
The finite element implementation of the interface model was established.
Within the homogenization framework, we derived semi-analytical expressions for the overall thermal conductivity of fiber- and particle-reinforced composites.
The results show an excellent agreement between the analytical estimates and the FEM simulations.
The overall response of the extended general interface model spans through all the values between the lower and upper bounds due to the LC and HC models, respectively.
Furthermore, by comparing our model to composites with graded interphases, we quantified the significance of the weighting factor in effectively capturing interphase effects.
Simply speaking, the weighting factor assigns more contribution to the side on which the more conductive interphase layers are located.
Overall, the proposed framework provides a versatile and physically consistent tool for modeling interfacial phenomena in heterogeneous materials.
Beyond offering a unifying perspective on existing interface models, it opens new possibilities for designing and optimizing composite materials by tailoring interphase characteristics to desired macroscopic properties.


\section*{Acknowledgement}

Paul Steinmann and Soheil Firooz gratefully acknowledge the support by the Deutsche Forschungsgemeinschaft (DFG, German Research Foundation) project number 460333672 – CRC 1540 Exploring Brain Mechanics (subproject C01).
Ali Javili gratefully acknowledges the support from The Scientific and Technological Research Council of Türkiye (T\"UB\.ITAK), grant number 224M144.


\appendix
\addtocontents{toc}{\protect\setcounter{tocdepth}{0}}

\section{Interface temperature coefficients in the Galerkin--Fourier composite assemblage methods}\label{sec:app_0}

This appendix collects the explicit expressions for the coefficients appearing in the interface temperature representation for both the GF-CCA and the GF-CSA methods.
Our MATLAB implementations are also provided, with inputs
\texttt{K1}, \texttt{K2}, \texttt{Ks}, \texttt{m}, \texttt{a}, \texttt{f}, \texttt{r}, \texttt{T0}, and \texttt{beta}
corresponding to $K^{(1)}$, $K^{(2)}$, $\o{K}$, $\o{m}$, $\o{\alpha}$, $f$, $r_1$, $\Theta_{\rm ref}$, and $\beta$, respectively.
The code solves Eqs.~\eqref{eq:ccaN2-normal-system} and~\eqref{eq:ccaN2-tangential-system2} and evaluates Eq.~\eqref{eq:ccaN2-keff2}.
To keep the code compact, the coefficient expansions are evaluated in their equivalent factored form using the normalized interface temperature increment $H=[\o{\Theta}-\Theta_{\rm ref}]/[\beta r_1]$.
The inputs should naturally satisfy $0<f<1$, $r_1>0$, $K^{(1)},K^{(2)}>0$, $\o{K},\o{m}\geq0$, $0\leq\o{\alpha}\leq1$, $\Theta_{\rm ref}>0$, and $\beta\neq0$.

\subsection{Coefficients for the GF-CCA}\label{sec:app_1}
\begin{equation}
\begin{aligned}
&N_0
=
\Theta_{\rm ref}^{2}
+
\f{\lb\beta r_1\rb^2}{2}
\Big[
A_1\lb C_1+D_1\rb
+
A_2\lb C_2+D_2\rb
\Big]\,,
\\
&N_1
=
\Theta_{\rm ref}\beta r_1
\lb
A_1+C_1+D_1
\rb
+
\f{\lb\beta r_1\rb^2}{2}
\Big[
A_1\lb C_2+D_2\rb
+
A_2\lb C_1+D_1\rb
\Big]\,,
\\
&N_2
=
\Theta_{\rm ref}\beta r_1
\lb
A_2+C_2+D_2
\rb
+
\f{\lb\beta r_1\rb^2}{2}
A_1\lb C_1+D_1\rb\,,
\\
&N_3
=
\f{1}{2}
\lb\beta r_1\rb^2
\Big[
A_1\lb C_2+D_2\rb
+
A_2\lb C_1+D_1\rb
\Big]\,,
\\
&N_4
=
\f{1}{2}
\lb\beta r_1\rb^2
A_2\lb C_2+D_2\rb\,,
\\
&M_0
=
\Theta_{\rm ref}\,,
\\
&M_1
=
\beta r_1
\Big[
\o{\alpha} A_1
+
\lb 1-\o{\alpha}\rb \lb C_1+D_1\rb
\Big]\,,
\\
&M_2
=
\beta r_1
\Big[
\o{\alpha} A_2
+
\lb 1-\o{\alpha}\rb \lb C_2+D_2\rb
\Big]\,.
\end{aligned}
\label{eq:ccaN2-harmonic-coefficients}
\end{equation}

\subsection{Coefficients for the GF-CSA}\label{sec:app_2}

The corresponding spherical MATLAB implementation solves Eqs.~\eqref{eq:CSA-N2-normal-system} and~\eqref{eq:CSA-N2-tangential-system2} and evaluates Eq.~\eqref{eq:CSA-N2-keff}.
The substitution $z=\cos\theta$ is used for the Legendre projections.

\begin{equation}
\begin{aligned}
N_0
&=
\Theta_{\text{ref}}^2
+
\lb \beta r_1 \rb^2
\lb
\f{1}{3}A_1\lb C_1+D_1 \rb 
+
\f{1}{5}A_2\lb C_2+D_2 \rb 
\rb\,,
\\
N_1
&=
\Theta_{\text{ref}}\beta r_1
\lb
A_1+C_1+D_1
\rb
+
\f{2}{5}\lb \beta r_1 \rb^2
\lb
A_1\lb C_2+D_2 \rb 
+
A_2\lb C_1+D_1 \rb 
\rb\,,
\\
N_2
&=
\Theta_{\text{ref}}\beta r_1
\lb
A_2+C_2+D_2
\rb
+
\lb \beta r_1 \rb^2
\lb
\f{2}{3}A_1\lb C_1+D_1 \rb 
+
\f{2}{7}A_2\lb C_2+D_2 \rb 
\rb\,,
\\
N_3
&=
\f{3}{5}\lb \beta r_1 \rb^2
\Big[
A_1\lb C_2+D_2 \rb 
+
A_2\lb C_1+D_1 \rb 
\Big]\,,
\\
N_4
&=
\f{18}{35}\lb \beta r_1 \rb^2
A_2\lb C_2+D_2 \rb \,,
\\
M_0
&=
\Theta_{\text{ref}}\,,
\\
M_1
&=
\beta r_1
\Big[
\o{\alpha}A_1
+
\lb 1-\o{\alpha} \rb \lb C_1+D_1 \rb 
\Big]\,,
\\
M_2
&=
\beta r_1
\Big[
\o{\alpha}A_2
+
\lb 1-\o{\alpha} \rb \lb C_2+D_2 \rb 
\Big]\,.
\end{aligned}
\label{eq:CSA-N2-harmonic-coefficients}
\end{equation}

\subsection{MATLAB Implementation of the Extended General Interface Model}\label{sec:app_3}
This appendix provides the MATLAB implementations used to obtain the two- and three-dimensional analytical estimates presented in this work. The codes correspond to the analytical formulations developed for the cylindrical and spherical representative volume elements and are included to facilitate reproducibility of the reported results.

\begin{spacing}{1}
\begin{lstlisting}[
    basicstyle=\ttfamily\footnotesize,
    upquote=true,
    breaklines=true,
    breakatwhitespace=false
]
%% Effective conductivity: extended general interface (2D or 3D)
% Author: Soheil Firooz
% Two-mode Fourier/Legendre Galerkin solution of a coated-cell model.
% Edit only this input section. Requires base MATLAB only.
clear; clc

PD = 2;                 % 2: circular inclusion; 3: spherical inclusion
f = 0.5;                % inclusion area/volume fraction (0 < f < 1)
alpha = 0.7;            % inclusion-side coldness weight (0 <= alpha <= 1)
Kinc = 1;               % inclusion bulk conductivity
Kmat = 1;               % matrix bulk conductivity
Kt = 100;               % tangential interface conductivity
Kn = 1;                 % normal interface conductivity (1/normal resistance)
L = 0.01;               % side length of equal-area/volume RVE
Theta0 = 0.01;          % positive reference temperature
DeltaTheta = 1e-3;      % imposed temperature difference across the RVE

Keff = calculateKeff(PD,f,alpha,Kinc,Kmat,Kt,Kn,L,Theta0,DeltaTheta);
fprintf('Keff = %.15g\n',Keff);

function Keff = calculateKeff(d,f,a,K1,K2,Kt,Kn,L,T0,dT)
    % The minus side is the inclusion and the plus side is the matrix.
    % 1/Tint = a/Tminus + (1-a)/Tplus.
    if ~ismember(d,[2 3]) || f<=0 || f>=1 || a<0 || a>1 || ...
            min([K1,K2,Kn,L,T0])<=0 || Kt<0 || dT==0
        error('Invalid inputs: check dimension, fractions, and positive properties.');
    end
    N = 2;                              % two angular modes
    r2 = (gamma(d/2+1)/pi^(d/2))^(1/d)*L;
    r = r2*f^(1/d);                    % inclusion radius
    beta = dT/L;

    % Angular basis and its first and second theta derivatives.
    if d==2
        M = 512;
        th = 2*pi*((0:M-1)'+0.5)/M;
        P  = [cos(th), cos(2*th)];
        Pt = [-sin(th), -2*sin(2*th)];
        Ptt = [-cos(th), -4*cos(2*th)];
        w = ones(M,1)*2/M;             % Fourier cosine projection
    else
        M = 128;
        j = (1:M-1)'; b = j./sqrt(4*j.^2-1);
        [V,D] = eig(diag(b,1)+diag(b,-1));
        [mu,idx] = sort(diag(D));
        w = 2*(V(1,idx)'.^2);           % Gauss-Legendre weights
        s = sqrt(1-mu.^2);
        P = [mu, (3*mu.^2-1)/2];
        Pt = [-s, -3*mu.*s];
        Ptt = [-mu, 3*(s.^2-mu.^2)];
        w = w.*[3/2,5/2];              % implicit mode-dependent projection
    end

    % First-mode bulk solution is used only to initialize the nonlinear solve.
    if d==2
        den = K1*(1-f)+K2*(1+f);
        A1 = 2*K2/den;
        B1 = (K1+K2)/den;
    else
        den = K1*(1-f)+K2*(2+f);
        A1 = 3*K2/den;
        B1 = (K1+2*K2)/den;
    end
    x = [A1;0;B1;0];              % inclusion and matrix growing-mode coefficients
    m = 1/Kn;
    sN = max(abs(beta)*(r+m*max(K1,K2)),1e-14);
    sT = max(abs(beta)*(max(K1,K2)+Kt/r),1e-14);
    Rfun = @(z) residual(z,d,f,a,K1,K2,Kt,m,r,beta,T0,P,Pt,Ptt,w,sN,sT);

    % Damped Newton method with a central-difference Jacobian.
    tol = 1e-10;
    for it = 1:80
        R = Rfun(x);
        if norm(R,inf)<tol, break; end
        J = zeros(4);
        for k = 1:4
            h = 1e-6*max(1,abs(x(k)));
            e = zeros(4,1); e(k) = h;
            J(:,k) = (Rfun(x+e)-Rfun(x-e))/(2*h);
        end
        if rcond(J)<1e-12, step = -pinv(J)*R;
        else, step = -J\R; end
        lambda = 1;
        for ls = 1:16
            trial = x+lambda*step;
            if all(isfinite(Rfun(trial))) && norm(Rfun(trial),inf)<norm(R,inf)
                break
            end
            lambda = lambda/2;
        end
        if ls==16 && norm(Rfun(trial),inf)>=norm(R,inf)
            trial = x+1e-3*step;
        end
        x = trial;
    end
    if norm(Rfun(x),inf)>=tol
        error('Nonlinear solve did not converge. Check inputs and temperature positivity.');
    end

    % Outer Dirichlet condition eliminates the decaying matrix modes.
    B = x(3:4);
    C = [(1-B(1))/f; -B(2)/f^(1+2/d)];
    Keff = K2*(B(1)-(d-1)*f*C(1));
end

function R = residual(x,d,f,a,K1,K2,Kt,m,r,beta,T0,P,Pt,Ptt,w,sN,sT)
    n = (1:2)';
    A = x(1:2); B = x(3:4);
    C = [(1-B(1))/f; -B(2)/f^(1+2/d)];
    V = B+C;

    % Temperature traces and tangential derivatives at the interface.
    tm = T0+beta*r*(P*A);
    tp = T0+beta*r*(P*V);
    tm_t = beta*r*(Pt*A);        tp_t = beta*r*(Pt*V);
    tm_tt = beta*r*(Ptt*A);      tp_tt = beta*r*(Ptt*V);
    qm = -K1*beta*(P*(n.*A));
    qp = -K2*beta*(P*(n.*B-(n+d-2).*C));

    % Harmonic interfacial temperature (weighted coldness average).
    D = a*tm+(1-a)*tp;
    if any(abs(D)<1e-12*max([ones(size(D)),abs(tm),abs(tp)],[],2))
        R = 1e8*ones(4,1); return
    end
    Hm = (1-a)*tp.^2./D.^2;
    Hp = a*tm.^2./D.^2;
    Hmm = -2*a*(1-a)*tp.^2./D.^3;
    Hpp = -2*a*(1-a)*tm.^2./D.^3;
    Hmp = 2*a*(1-a)*tm.*tp./D.^3;
    Tt = Hm.*tm_t+Hp.*tp_t;
    Ttt = Hm.*tm_tt+Hp.*tp_tt+Hmm.*tm_t.^2+ ...
          2*Hmp.*tm_t.*tp_t+Hpp.*tp_t.^2;

    % Normal jump and tangential surface-balance conditions.
    Lap = Ttt;
    if d==3
        mu = P(:,1);
        Lap = Lap+mu./sqrt(1-mu.^2).*Tt;
    end
    Fn = tp-tm+m*((1-a)*qp+a*qm);
    Ft = qp-qm-Kt/r^2*Lap;
    if d==2
        R = [P'*(w.*Fn)/sN;P'*(w.*Ft)/sT];
    else
        R = [sum(w.*(P.*Fn),1)'/sN;sum(w.*(P.*Ft),1)'/sT];
    end
end
\end{lstlisting}
\end{spacing}


\bibliography{library}


\end{document}